\documentclass{aa}
\usepackage{multirow}
\usepackage{amsmath}
\usepackage[colorlinks,linkcolor=blue,anchorcolor=blue,citecolor=blue]{hyperref}
\usepackage{graphicx} 
\usepackage{lscape}
\usepackage{longtable}
\usepackage{txfonts}
\usepackage{natbib}
\usepackage{color}

\begin{document}

\title{The SRG/eROSITA All-Sky Survey} 
\subtitle{First catalog of superclusters in the western Galactic hemisphere \thanks{Tables 1, 2, A.1, and A.2 are only available in electronic form at the CDS via anonymous ftp to cdsarc.u-strasbg.fr (130.79.128.5) or via http://cdsweb.u-strasbg.fr/cgi-bin/qcat?J/A+A/.}}

\author{
A.~Liu\inst{1}\thanks{e-mail: \href{mailto:liuang@mpe.mpg.de}{\tt liuang@mpe.mpg.de}},
E.~Bulbul\inst{1},
M.~Kluge\inst{1},
V.~Ghirardini\inst{1},
X.~Zhang\inst{1},
J.~S.~Sanders\inst{1},
E.~Artis\inst{1},
Y.~E.~Bahar\inst{1},
F.~Balzer\inst{1},
M.~Br\"uggen\inst{2},
N.~Clerc\inst{3},
J.~Comparat\inst{1},
C.~Garrel\inst{1},
E.~Gatuzz\inst{1},
S.~Grandis\inst{4},
G.~Lamer\inst{5},
A.~Merloni\inst{1},
K.~Migkas\inst{6,7},
K.~Nandra\inst{1},
P.~Predehl\inst{1},
M.~E.~Ramos-Ceja\inst{1},
T.~H.~Reiprich\inst{7},
R.~Seppi\inst{1},
S.~Zelmer\inst{1}
}
\institute{
\inst{1}{Max Planck Institute for Extraterrestrial Physics, Giessenbachstrasse 1, 85748 Garching, Germany}\\
\inst{2}{Hamburger Sternwarte, Universität Hamburg, Gojenbergsweg 112, 21029 Hamburg, Germany}\\
\inst{3}{IRAP, Universitt’e de Toulouse, CNRS, UPS, CNES, F-31028 Toulouse, France}\\
\inst{4}{Institute for Astro- and Particle Physics, University of Innsbruck, Technikerstr. 25, 6020 Innsbruck, Austria}\\
\inst{5}{Leibniz-Institut für Astrophysik Potsdam (AIP), An der Sternwarte 16, 14482 Potsdam, Germany}\\
\inst{6}{Leiden Observatory, Leiden University, PO Box 9513, 2300 RA Leiden, the Netherlands}\\
\inst{7}{Argelander-Institut für Astronomie, Universität Bonn, Auf dem Hügel 71, 53121 Bonn, Germany}
}

\titlerunning{eRASS1 supercluster catalog}
\authorrunning{Liu et al.}

\abstract
{Superclusters of galaxies mark the large-scale overdense regions in the Universe. Superclusters provide an ideal environment to study structure formation and to search for the emission of the intergalactic medium such as cosmic filaments and WHIM. In this work, we present the largest-to-date catalog of X-ray-selected superclusters identified in the first SRG/eROSITA All-Sky Survey (eRASS1). By applying the Friends-of-Friends (FoF) method on the galaxy clusters detected in eRASS1, we identified 1338 supercluster systems in the western Galactic hemisphere up to redshift 0.8, including 818 cluster pairs and 520 rich superclusters with $\ge 3$ members. The most massive and richest supercluster system is the Shapley supercluster at redshift 0.05 with 45 members and a total mass of $2.58\pm0.51 \times10^{16} M_{\odot}$. The most extensive system has a projected length of 127~Mpc. The sizes of the superclusters we identified in this work are comparable to the structures found with galaxy survey data. We also found a good association between the eRASS1 superclusters and the large-scale structures formed by optical galaxies. We note that 3948 clusters, corresponding to $45\%$ of the cluster sample, were identified as supercluster members. The reliability of each supercluster was estimated by considering the uncertainties in the redshifts of the galaxy clusters and the peculiar velocities of clusters. Furthermore, 63\% of the systems have a reliability larger than 0.7. The eRASS1 supercluster catalog provided in this work represents the most extensive sample of superclusters selected in the X-ray band in terms of the unprecedented sample volume, sky coverage, redshift range, the availability of X-ray properties, and the well-understood selection function of the parent cluster sample, which enables direct comparisons with numerical simulations. This legacy catalog will greatly advance our understanding of superclusters and the cosmic large-scale structure. }


\maketitle

\section{Introduction}

In the hierarchical structure formation picture, galaxy clusters emerge from the rarest and highest density peaks of initial fluctuations and become gravitationally bound after going through several merging and accretion processes \citep[see][for a review]{kravtsov2012}. While clusters are widely used to study cosmology and structure formation, their spatial distribution also traces the cosmic overdense regions on a larger scale, such as superclusters. As components of the cosmic web, clusters and superclusters mark the density peaks in the large-scale structure. They are connected by other elements, such as filaments, sheets, and walls, and are separated by low-density regions, such as cosmic voids.

Superclusters are groups of galaxy clusters consisting of more than one member cluster. Superclusters will not necessarily collapse in the future as they are not gravitationally bound, and the physical connections between their member clusters are relatively weak compared to virialized systems. Except for some cases of merging clusters, in most of the supercluster systems, their members have not yet interacted with other members and lie beyond the virial radius of each other. These features make it difficult to define and identify superclusters in observations quantitatively. Although the concept of a supercluster was introduced about seventy years ago \citep{1953deVaucouleurs}, a precise and widely accepted definition of superclusters is still absent. \citet{2014Tully} and \citet{2019Einasto} proposed to define superclusters based on their dynamic effect on the cosmic environment, calling them ``basins of attraction'' and ``cocoons;'' \citet{2015Chon} suggested to call the superclusters that will survive the cosmic expansion and eventually collapse in the future ``superstes-clusters,'' to distinguish them from traditional superclusters. In fact, some traditional superclusters, such as Laniakea \citep{2014Tully}, are essentially compilations of several smaller superclusters, and the whole system will disperse in the future \citep{2015Chon}. 

Despite the difficulties in defining superclusters, there have been numerous works aiming at finding superclusters based on optical galaxy surveys or optical-selected cluster catalogs for decades \citep[see, e.g.,][for some early attempts]{1961Abell,1993Zucca,1994Einasto}. Most of the works where superclusters have been identified with galaxy catalogs used the method of galaxy density field. They computed the density of galaxies across the sky and searched for peaks of galaxy densities. For example, \citet{2007Einasto} found 543 superclusters with the Two-degree-Field Galaxy Redshift Survey data by selecting the peaks in the galaxy density field. Using the Sloan Digital Sky Survey Data Release 7 galaxy catalog, \citet{2012Liivamagi} constructed a set of supercluster catalogs by searching for regions with densities over a selected threshold. Using a fixed density threshold and an adaptive local density threshold, they found 982 and 1313 superclusters, respectively. On the other hand, the works where superclusters were identified directly from cluster catalogs prefer the Friends-of-Friends (FoF) method, mostly due to the difficulties in precisely computing the density field of clusters as they are much rarer than galaxies. For example, \citet{2023Sankhyayan} identified 662 superclusters in the redshift range [0.05, 0.42] by applying a modified FoF method on the Wen-Han-Liu (WHL) SDSS cluster catalog \citep{2012Wen}. In summary, the number of known optical superclusters up to now is on the order of 10$^3$.

Compared to optical cluster surveys, X-ray cluster surveys have the advantage in terms of sample purity, as the former often suffers from projection effects \citep[e.g.,][]{2019Costanzi,2021Myles}. As clusters are more highly biased than galaxies \citep[see, e.g.,][]{Seppi2023}, fewer of them are needed to significantly trace the same underlying structure.
Thanks to the availability of large-area X-ray surveys, in the past decades, there have been an increasing number of efforts in detecting superclusters directly using X-ray cluster samples \citep{2001Einasto,2013Chon,2018Adami,Boehringer2021,2022Liu}. The first X-ray flux-limited supercluster sample was constructed by \citet{2013Chon}, with the extended ROSAT-ESO Flux-Limited X-ray (REFLEX II) galaxy cluster sample. They found 164 superclusters below redshift 0.4. In recent years, supercluster catalogs have become important products of X-ray cluster surveys. In the XXL survey, \citet{2018Adami} detected 35 superclusters with $N_{\rm cl} \ge 3$, and 39 cluster pairs, out to redshift 0.8. In the {\sl eROSITA} Final Equatorial-Depth Survey \citep[eFEDS,][]{2022Brunner}, \citet{2022Liu} detected 84 superclusters including 19 rich systems with $N_{\rm cl} \ge 4$. \citet{Boehringer2021} identified eight superclusters in the local Universe ($z\le 0.03$) from the Cosmic Large-Scale Structure in X-rays cluster survey, compiled from the X-ray clusters in the ROSAT All-Sky Survey.

Most of these works used the FoF algorithm to identify superclusters, where the linking length was determined based on a specific local overdensity ratio at a specific redshift with respect to the sample average. It should be noted, however, that the overdensity ratio $f$ can vary in different works \citep[e.g.,][]{2013Chon,Boehringer2021}, and the detection of superclusters is sensitive to the choice of $f$ \citep[see, e.g.,][]{2013Chon}. In the nearby Universe, since the X-ray cluster surveys are almost complete at high fluxes, one can also use the cluster X-ray luminosity function (XLF) to compute the local overdensity with respect to the cosmic average \citep[e.g.,][]{Boehringer2021}, which would ideally give the same results as using the sample average. Other methods, such as the Voronoi tessellation, have also been used in the identification of superclusters and are found to give similar results as FoF \citep{2018Adami}. Recently, another novel technique based on the dendrogram of galaxies was developed and applied to the identification of clustering patterns of stars and galaxies \citep{Liu2018b,2022Yu}, which also has promising prospects of application in supercluster detection.

\begin{figure}
\begin{center}
\includegraphics[width=0.49\textwidth, trim=20 40 55 10, clip]{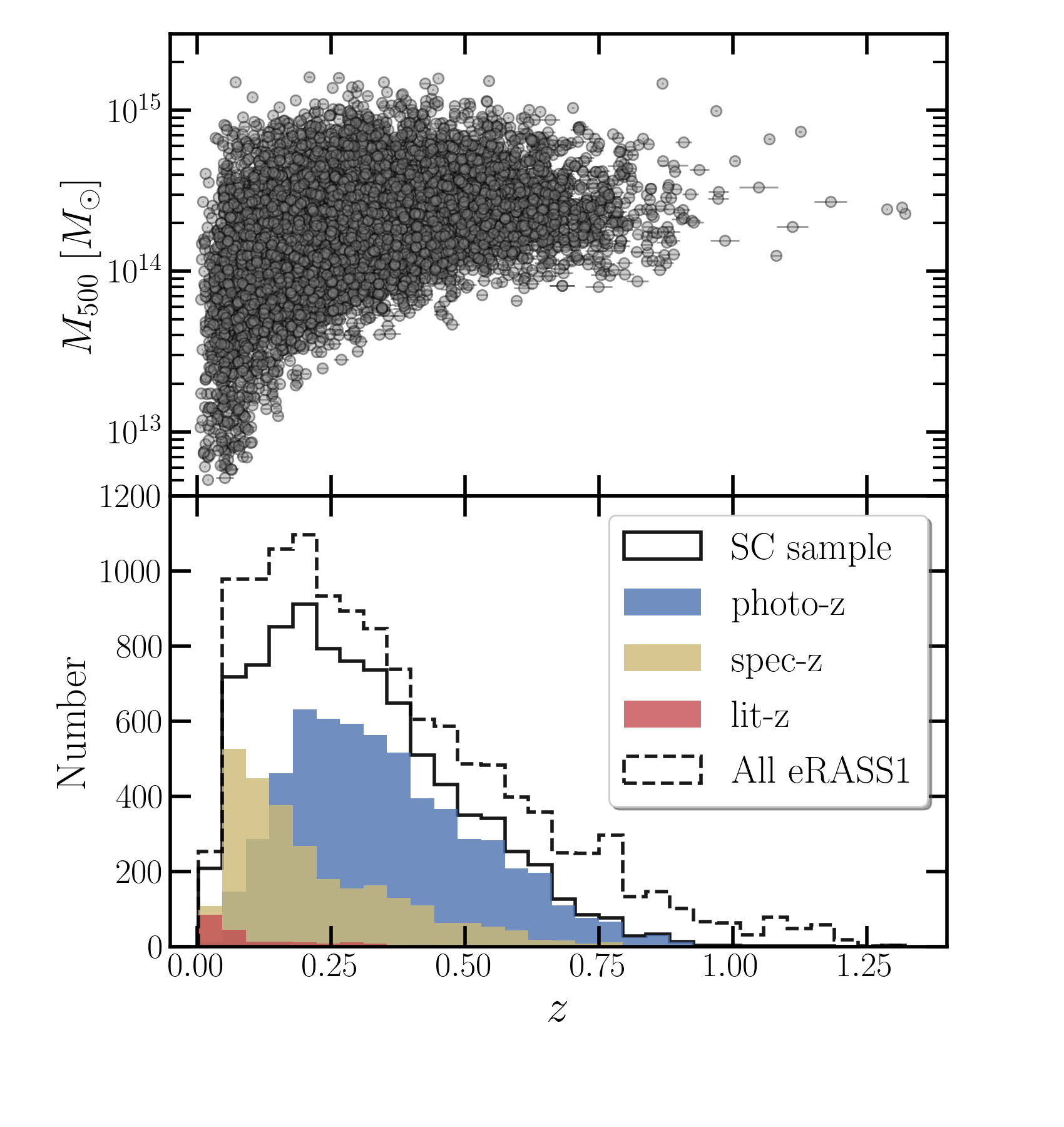}
\caption{  \label{sample} Mass and redshift distribution of the cluster sample used for supercluster identification. Blue, yellow, and red histograms indicate photometric, spectroscopic, and literature redshifts, respectively. The black dashed line shows the redshift distribution of the overall eRASS1 cluster sample. }
\end{center}
\end{figure}

\begin{figure*}
\begin{center}
\includegraphics[width=0.495\textwidth, trim=0 25 0 70, clip]{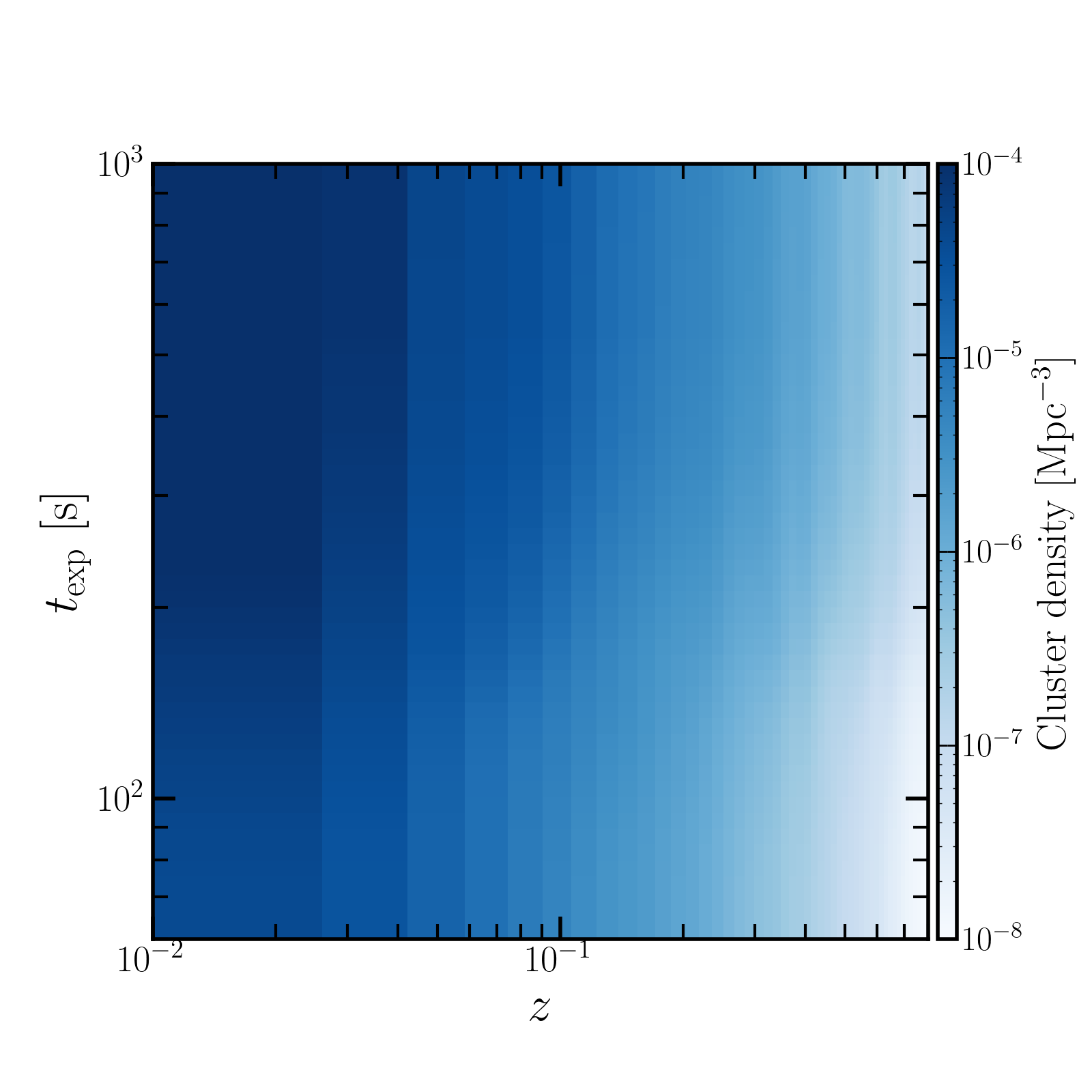}
\includegraphics[width=0.495\textwidth, trim=0 25 0 70, clip]{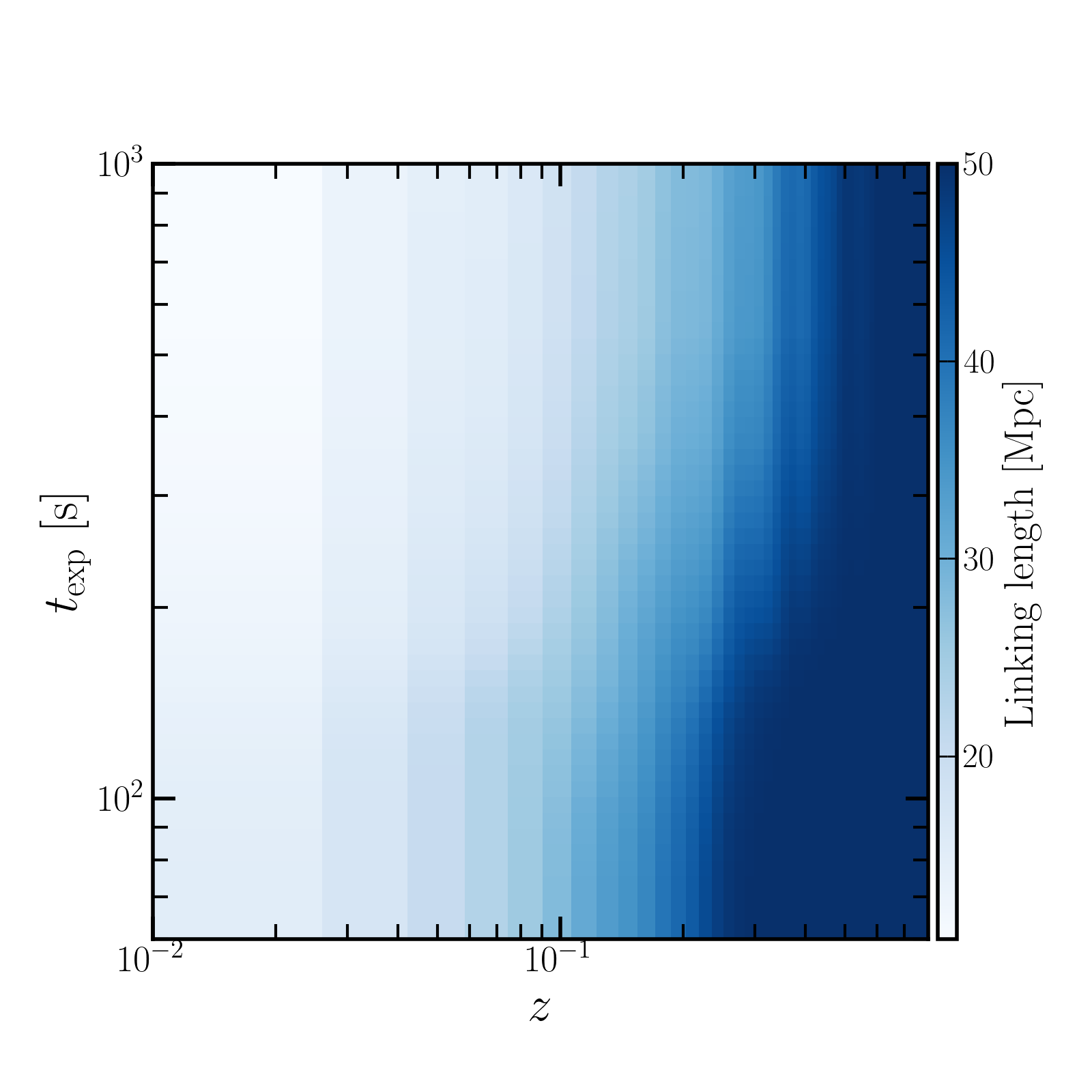}
\caption{\label{lk} Cluster number density ($N/V$, left panel) and linking length ($l$, right panel) plotted as functions of both redshift and exposure time, computed using Eq.~\ref{eqlk}, with $f=10$. }
\end{center}
\end{figure*}

A large and representative sample of superclusters will advance the studies on both cosmology and astrophysics in many aspects. As the largest elements in the cosmic web, superclusters retain the history of the formation and evolution of the web. As ``clusters of clusters'', they are ideal laboratories to investigate the environmental effect on the evolution of galaxy clusters and galaxies. Superclusters can also play essential roles in the search for warm-hot intergalactic medium (WHIM) thanks to their association with cosmic structures. Moreover, nonthermal physics such as magnetic fields and cosmic rays can be studied in these dynamic systems.
The amount of attempts to directly use superclusters as cosmology probes is quite limited. This is owing mostly to the difficulties in precisely identifying superclusters, such as their centers, masses, and edges, and also to the sample volume: the total number of X-ray superclusters is on the order of 10$^2$ \citep{2001Einasto,2013Chon,2018Adami,2022Liu}. On the other hand, some well-known supercluster systems have already been used to study the large-scale structure, for example, mapping the structures in the nearby Universe \citep[e.g.,][]{Boehringer2021,Boehringer2021b}, and detecting or characterizing the inter-cluster filaments within supercluster systems \citep[e.g.,][]{2015Ursino,2016Bulbul,2021Reiprich,2021Ghirardini,2023Hoang}.

A precondition of finding superclusters is large samples of clusters with high sample purity and precise redshifts. These are available thanks to the recent large-area surveys in X-ray, Sunyaev-Zeldovich (SZ), and optical bands. In particular, the {\sl eROSITA} \citep[extended ROentgen Survey with an Imaging Telescope Array,][]{2021Predehl} X-ray telescope onboard Spectrum Roentgen Gamma (SRG) will detect about $10^5$ clusters and groups during its lifetime \citep{Merloni2012,2022Liu,2022Bulbul}, which will substantially increase the sample size of X-ray clusters. In eFEDS, 84 supercluster systems are detected in the $\sim140$ deg$^2$ survey area \citep{2022Liu}. Projected from the eFEDS results, the final {\sl eROSITA} All-Sky Survey (eRASS) is expected to detect thousands of superclusters in the western Galactic hemisphere. In particular, thanks to the unprecedented sensitivity in the soft X-ray band, {\sl eROSITA} will detect a large number of galaxy groups, which trace the large-scale structure even better than massive clusters owing to the overwhelming advantage in numbers. In the {\sl eROSITA} and eRASS era, we can trace the large-scale structure with comparable power to the optical galaxy surveys for the first time. The large sample of superclusters detected by {\sl eROSITA} will expand the study of superclusters from nearby Universe to higher redshifts ($z\approx 1$), from single targets to representative populations, and will greatly advance our understanding of superclusters and the related cosmological and astrophysical topics. 

In this work, we search for supercluster systems based on the galaxy cluster catalog from the first {\sl eROSITA} All-Sky Survey \citep[eRASS1,][]{Merloni2023,Bulbul2023,Kluge2023}. Leveraging this largest-ever X-ray galaxy cluster sample, we aim to present the biggest X-ray supercluster catalog to be used for further explorations of superclusters. The paper is organized as follows. In Sect.~\ref{sec:cluster}, we introduce the eRASS1 galaxy cluster sample used for supercluster detection. In Sect.~\ref{sec:detection}, we describe the identification of eRASS1 superclusters. In Sect.~\ref{sec:property}, we study the properties of the eRASS1 superclusters and their members. Our conclusions are summarized in Sect.~\ref{sec:conclusions}. Throughout this paper, we adopt the concordance $\Lambda$CDM cosmology with $\Omega_{\Lambda} =0.7$, $\Omega_{\mathrm m} =0.3$, and $H_0 = 70$~km~s$^{-1}$~Mpc$^{-1}$. However, we note that the exact choice of cosmological parameters does not affect the results significantly. Quoted error bars correspond to a 1$\sigma$ confidence level. To avoid confusion, we refer to the supercluster systems with two member clusters as ``cluster pairs'', and the systems with more than two member clusters as ``rich superclusters'', unless noted otherwise.

\section{Galaxy cluster sample}
\label{sec:cluster}

\subsection{eRASS1 galaxy cluster catalog}
\label{sec:cluster1}
The first {\sl eROSITA} All-Sky Survey was completed on June 11, 2020. In the western Galactic hemisphere\footnote{Defined as ($179.9442^\circ < l < 359.9442^\circ$)}, nearly 9.3$\times10^5$ X-ray sources are detected in the 0.2--2.3~keV energy range where {\sl eROSITA} is most sensitive \citep{Merloni2023}. Among them, about $2.7\times10^4$ are classified as extended sources, namely, galaxy cluster candidates, according to a simple cut on X-ray extent likelihood: $\mathcal{L}_{\rm ext}>3$ \citep{Bulbul2023}. 

\begin{figure*}
\begin{center}
\includegraphics[width=0.99\textwidth, trim=50 80 50 95, clip]{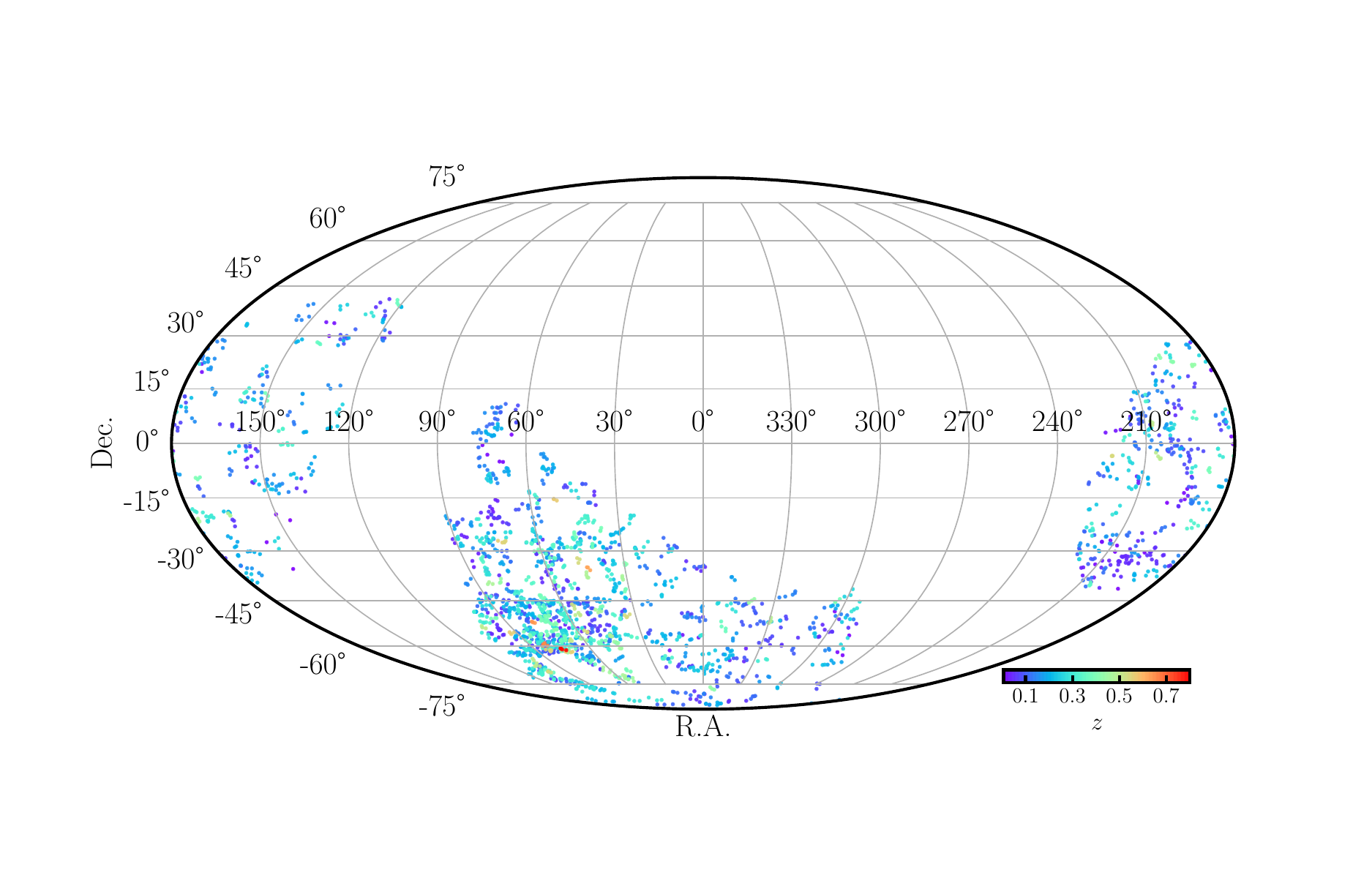}
\caption{\label{fig:scall} Distribution of the eRASS1 supercluster systems color-coded by redshift. Each point represents a supercluster member. For clarity purposes, only rich superclusters with $\ge 3$ member clusters are plotted. }
\end{center}
\end{figure*}

Optical identification of the cluster candidates is performed with {\tt eROMaPPer}, a highly parallelized version of the red-sequence-based cluster finder tool {\tt redMaPPer} \citep{2014Rykoff,2016Rykoff}. The public photometric data of the DESI Legacy Imaging Surveys DR9 and DR10 \citep{2019Dey} are used for the confirmation of clusters and the computation of photometric redshifts $z_{\lambda}$. Where possible, we also derive the spectroscopic redshifts of the eRASS1 clusters using publicly available galaxy spectroscopic redshifts. The final choice of cluster redshift {\tt BEST\_Z} is made as follows. If the cluster has at least three spectroscopic members, we determine its {\tt BEST\_Z} as $z_{\rm spec}$, which is the mean of the spectroscopic redshifts of the members. Otherwise, if the optical central galaxy has a spectroscopic redshift $z_{\rm spec,cg}$, we determine {\tt BEST\_Z} as $z_{\rm spec,cg}$. When neither of the above conditions is met, we adopt the photometric redshift $z_{\lambda}$ as the {\tt BEST\_Z}. Finally, for a few hundred clusters without spectroscopic or photometric redshifts, we adopt the literature redshifts $z_{\rm lit}$ by matching with public cluster catalogs using a matching radius of 2\arcmin. In summary, the eRASS1 galaxy cluster catalog contains 12247 optically confirmed clusters up to redshift 1.32. The details of the optical follow-up methodology are presented in \citet{Kluge2023}.

To obtain the X-ray properties of the eRASS1 clusters, a multi-band X-ray imaging analysis is performed for each cluster, using the tool {\tt MBProj2D} \citep{2018Sanders}. By forward-fitting cluster's X-ray images in seven bands from 0.3~keV to 7~keV, {\tt MBProj2D} provides the best-fit cluster physical model. Products for each cluster include the azimuthally-averaged electron density profile described in the form of the model in \citet{vikhlinin2006a}, a single global temperature under the isothermal assumption, and other derived quantities such as flux, luminosity, count rate, gas mass, etc. In particular, the above quantities are given as a function of radius and in multiple energy ranges \citep[see, e.g.,][for a recent application of {\tt MBProj2D}]{2023Liu}. Masses within $R_{500}$\footnote{$R_{500}$ is the radius within which the average matter density is 500 times the critical density at cluster’s redshift} of the clusters are computed based on the scaling relation between X-ray count rate, redshift, and mass, after calibrated with weak lensing shear signal \citep[see more details in][]{Grandis2023,Ghirardini2023}. The masses of eRASS1 clusters span a range $5\times 10^{12}M_{\odot} < M_{500} < 2\times 10^{15}M_{\odot}$.

A contamination probability estimator, $P_{\rm cont}$, is also computed for each eRASS1 cluster. The calculation of $P_{\rm cont}$ is based on a mixture model that takes into account the cluster's redshift, X-ray count rate, and optical richness. It gives the probability of a cluster to be a contaminant. According to the results of $P_{\rm cont}$ for each cluster, we estimate that around 1700 of the 12247 eRASS1 clusters are spurious. This is computed by $\sum P_{\rm cont}$. Thus the purity of the eRASS1 cluster sample, $\sum (1-P_{\rm cont})/N_{\rm cl}$, is about $86\%$. Most of the contaminants in the catalog are AGN, misclassified as extended sources due to the sizable PSF of {\sl eROSITA}. More details about the eRASS1 cluster catalog are provided in \citet{Bulbul2023} and \citet{Kluge2023}.

\begin{figure}
\begin{center}
\includegraphics[width=0.49\textwidth, trim=20 40 50 40, clip]{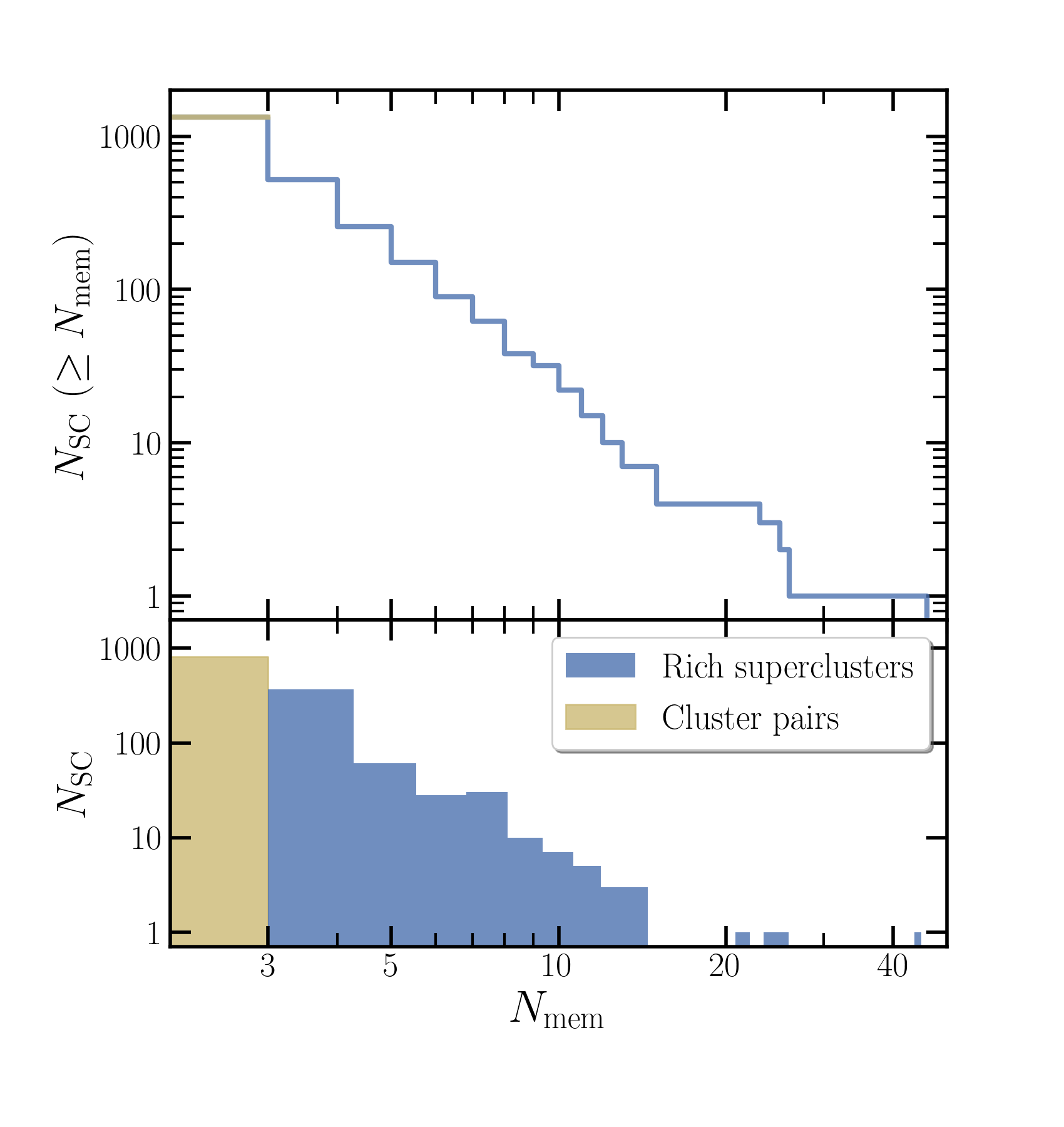}
\caption{\label{fig:multi} Multiplicity of the eRASS1 superclusters. Cluster pairs and superclusters (with $\ge 3$ member clusters) are plotted in yellow and blue, respectively. }
\end{center}
\end{figure}

\begin{figure*}
\begin{center}
\includegraphics[width=0.99\textwidth, trim=80 0 100 50, clip]{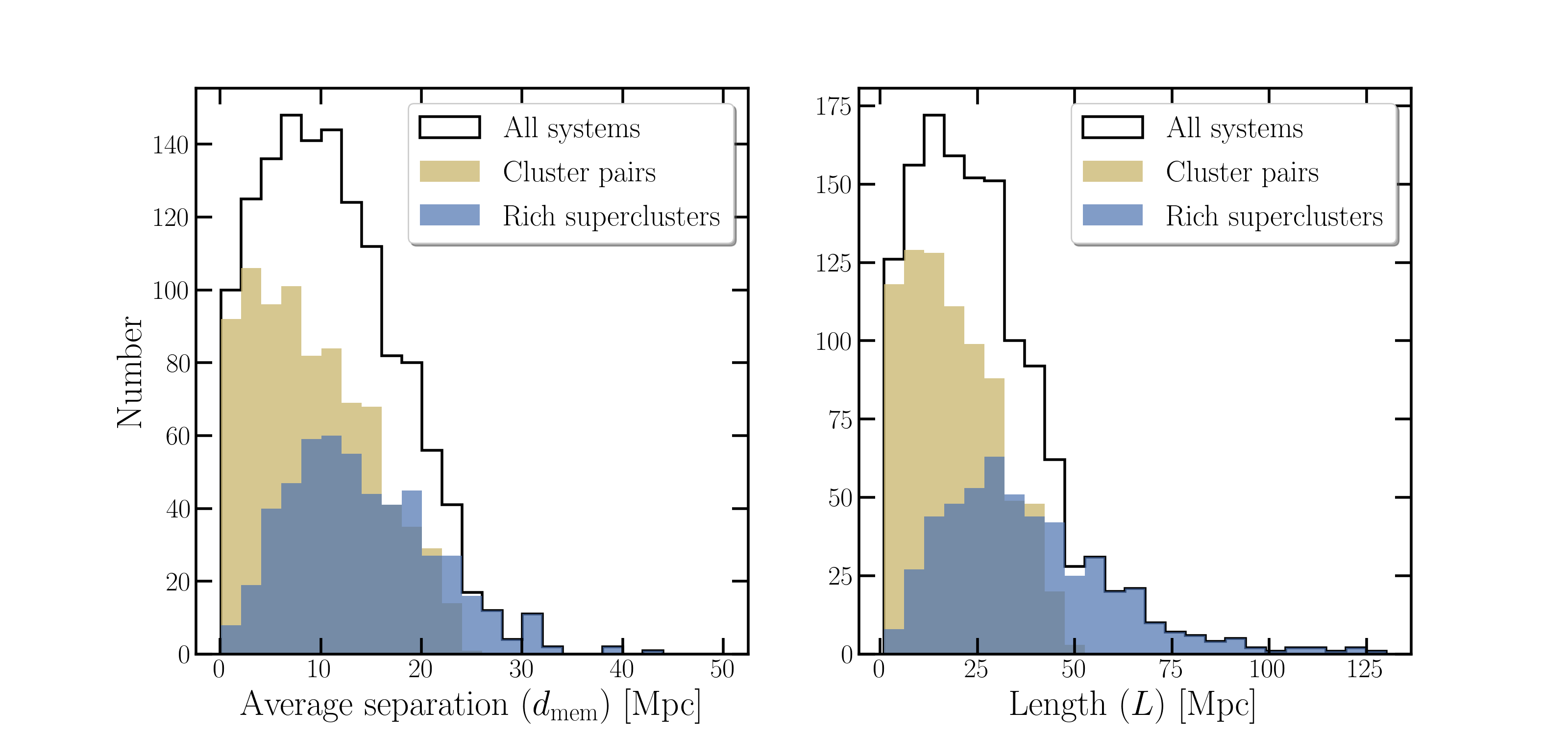}
\caption{\label{fig:dist} Average separation and total length of the superclusters. Both quantities are given in projected co-moving distance. Cluster pairs and rich superclusters are plotted in yellow and blue, respectively. }
\end{center}
\end{figure*}

\subsection{Sample selection}
\label{sec:cluster2}
We apply several additional selections on the primary eRASS1 cluster catalog published in \citet{Bulbul2023} to obtain a subsample of clusters with higher purity and more reliable redshifts, which is suitable for supercluster detection. Firstly, about $1900$ clusters with $P_{\rm cont}$ larger than 0.3 are ignored to enhance the purity of the sample. Secondly, since the detection of superclusters is sensitive to cluster redshifts, we exclude the clusters with unreliable redshifts. These include about $900$ cases when the errorbar in redshift is large ($\delta z/(1+z) > 0.02$) or when the photometric redshift exceeds the limiting redshift at cluster's sky position \citep[see][for more details]{Bulbul2023,Kluge2023}. Thirdly, about 300 clusters are found to be duplicates based on the optical data. Namely, more than 70\% of their members are also identified as members of another cluster, which has a higher extent likelihood \citep[see][and the optical follow-up catalog, for more information]{Kluge2023}. These clusters are also excluded from the sample. Fourthly, we exclude about $200$ clusters with less than 5 X-ray photons within $R_{500}$ to further reduce the contamination level. Finally, a few nearby clusters below redshift 0.005, such as Virgo, are not considered. With the above selections, we obtain a subsample of 8862 clusters in the redshift range of [0.0056, 1.32], with a purity of 96.4\% estimated from the $P_{\rm cont}$ of the remaining clusters in the subsample. 

In Fig.~\ref{sample}, we present the mass $M_{500}$ and redshift distribution of the cluster sample. Most high redshift clusters do not pass the selection criteria due to the larger error bar in their photometric redshifts. Among the 8862 clusters in the sample, 2758 have spectroscopic redshifts as {\tt BEST\_Z}, 5885 and 219 have photometric and literature redshifts respectively. The median redshift of the sample is 0.28, slightly lower than the overall eRASS1 cluster sample, where $z_{\rm median} \approx 0.31$. The loss of high redshift clusters will limit our supercluster identification to below $z\approx 1$.

\section{Supercluster identification}
\label{sec:detection}

\begin{figure*}
\begin{center}
\includegraphics[width=0.95\textwidth, trim=40 40 5 100, clip]{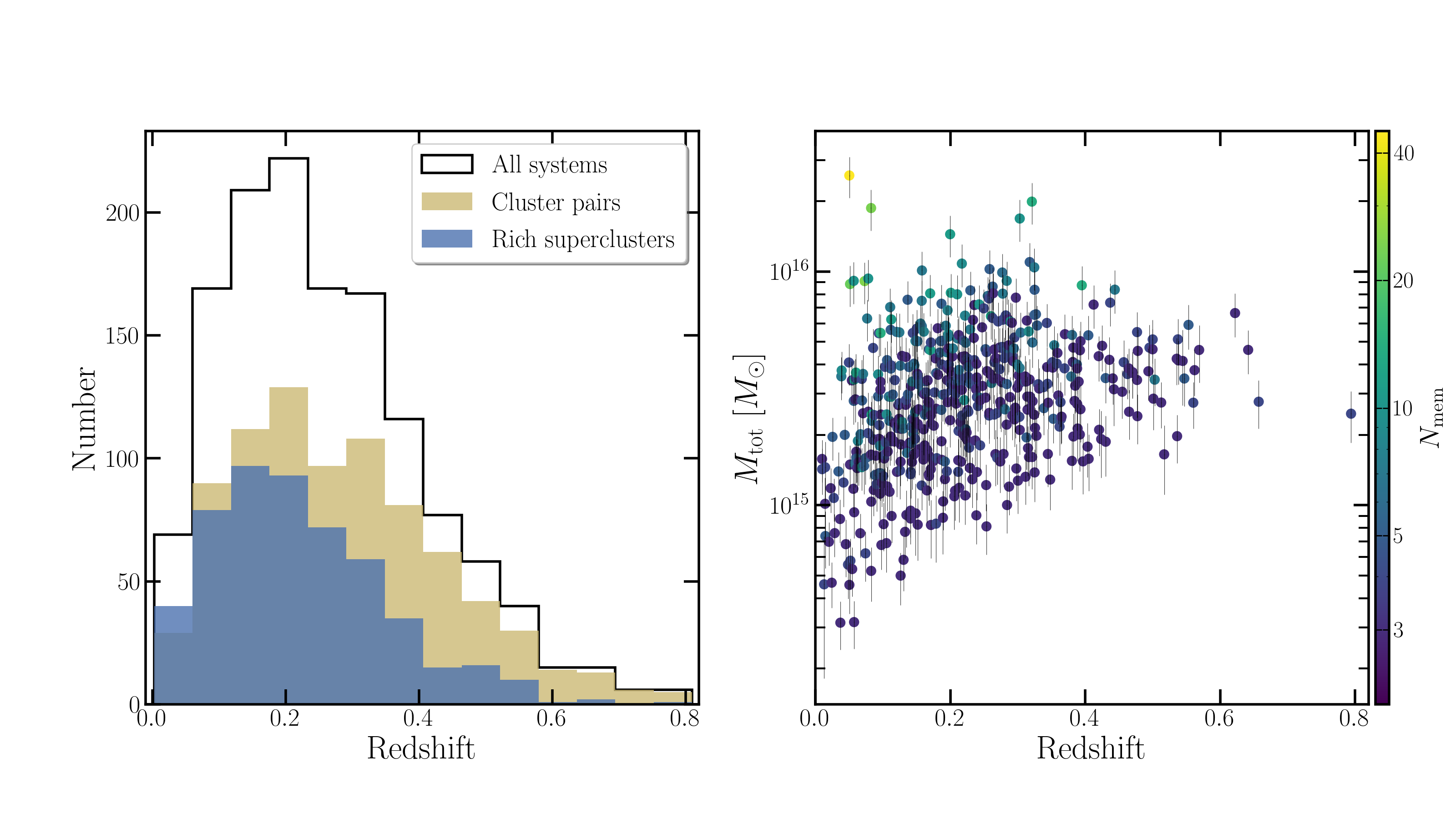}
\caption{\label{fig:scz} General properties of the supercluster systems identified in eRASS1. Left panel: redshift distribution of eRASS1 superclusters. Cluster pairs and superclusters (with $\ge 3$ member clusters) are plotted in yellow and blue, respectively. Right panel: supercluster total mass versus redshift. Color code denotes the number of member clusters. For clarity, only rich superclusters with $\ge 3$ members are shown in this plot. }
\end{center}
\end{figure*}

\begin{figure*}
\begin{center}
\includegraphics[width=0.495\textwidth, trim=0 40 40 60, clip]{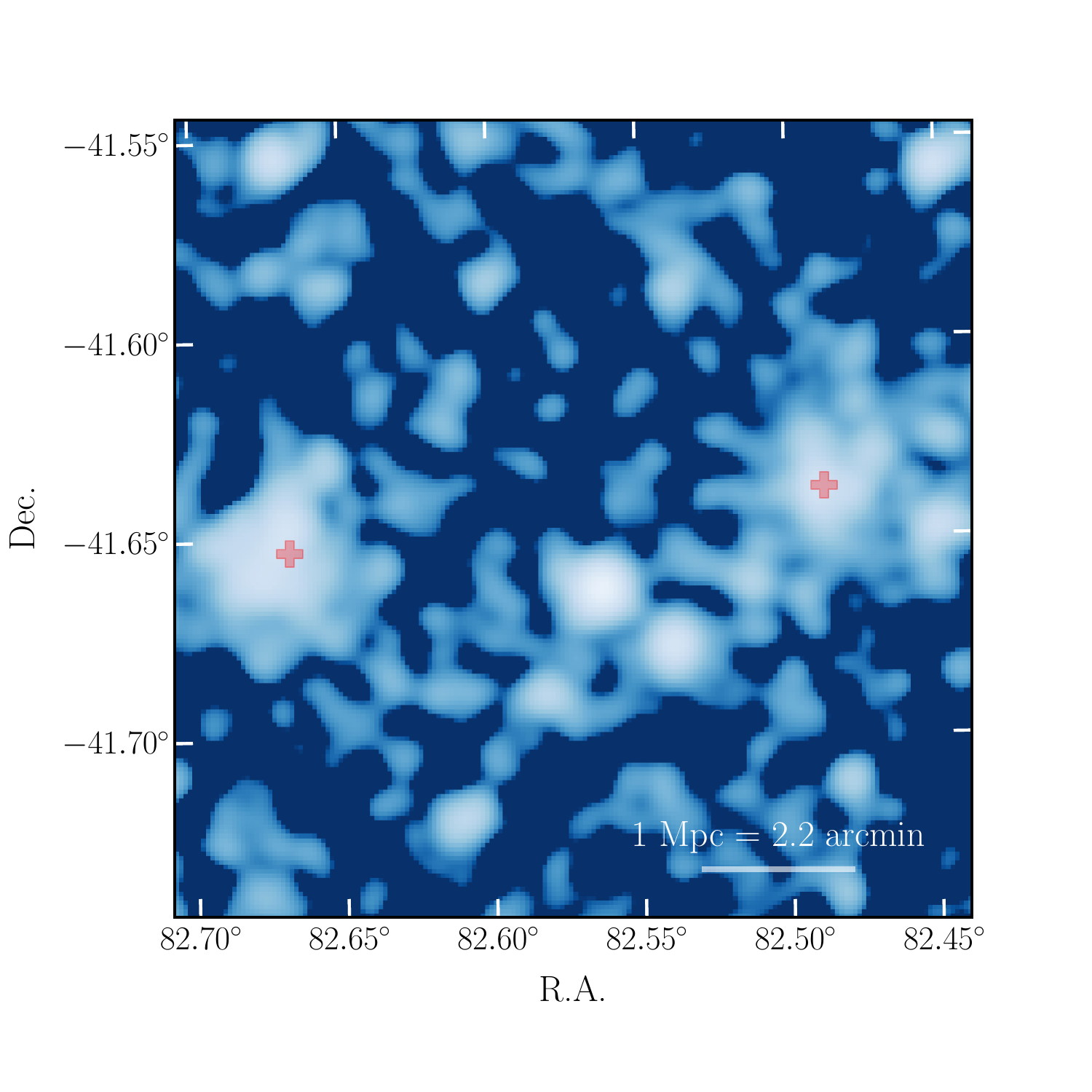}
\includegraphics[width=0.495\textwidth, trim=0 40 40 60, clip]{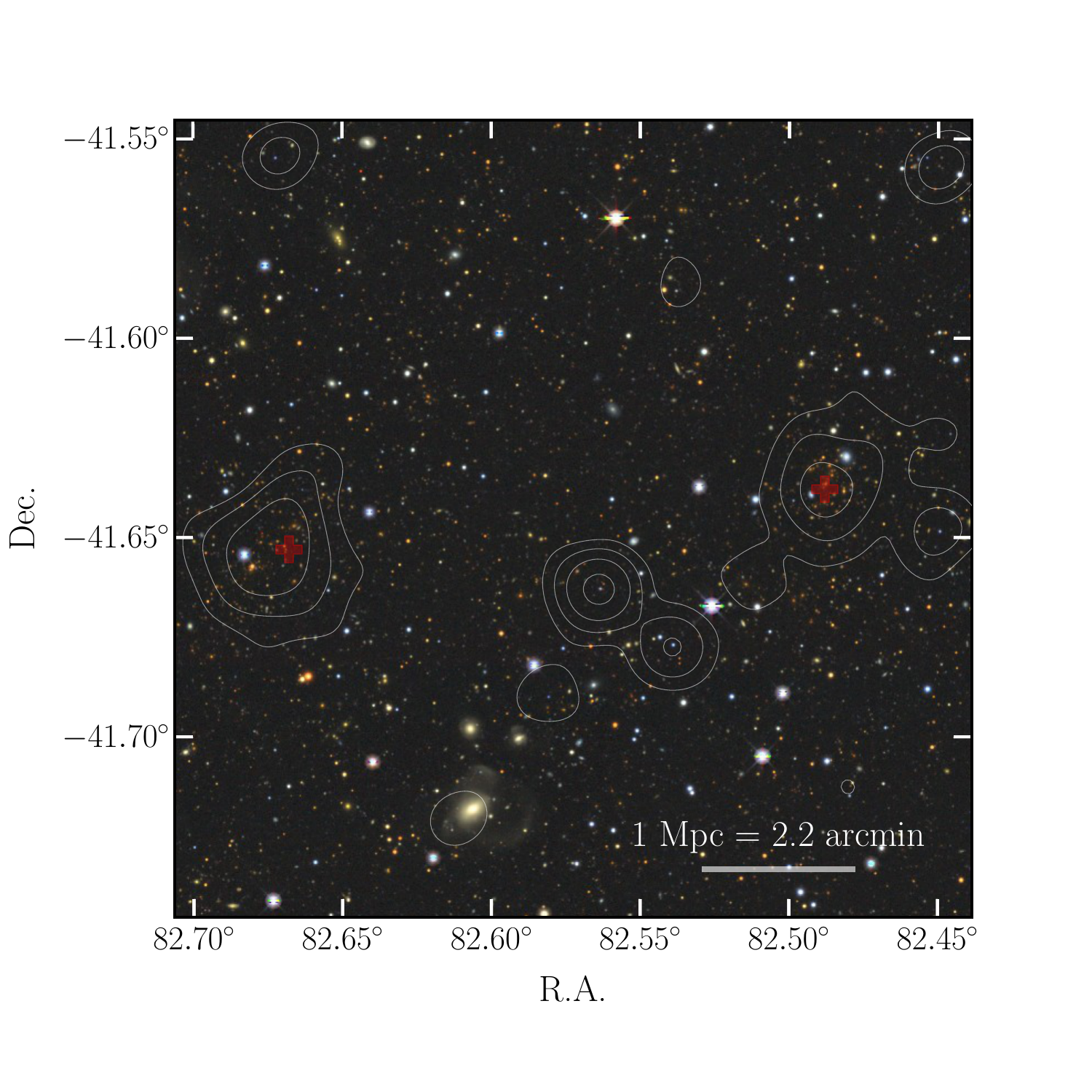}
\includegraphics[width=0.495\textwidth, trim=0 40 40 60, clip]{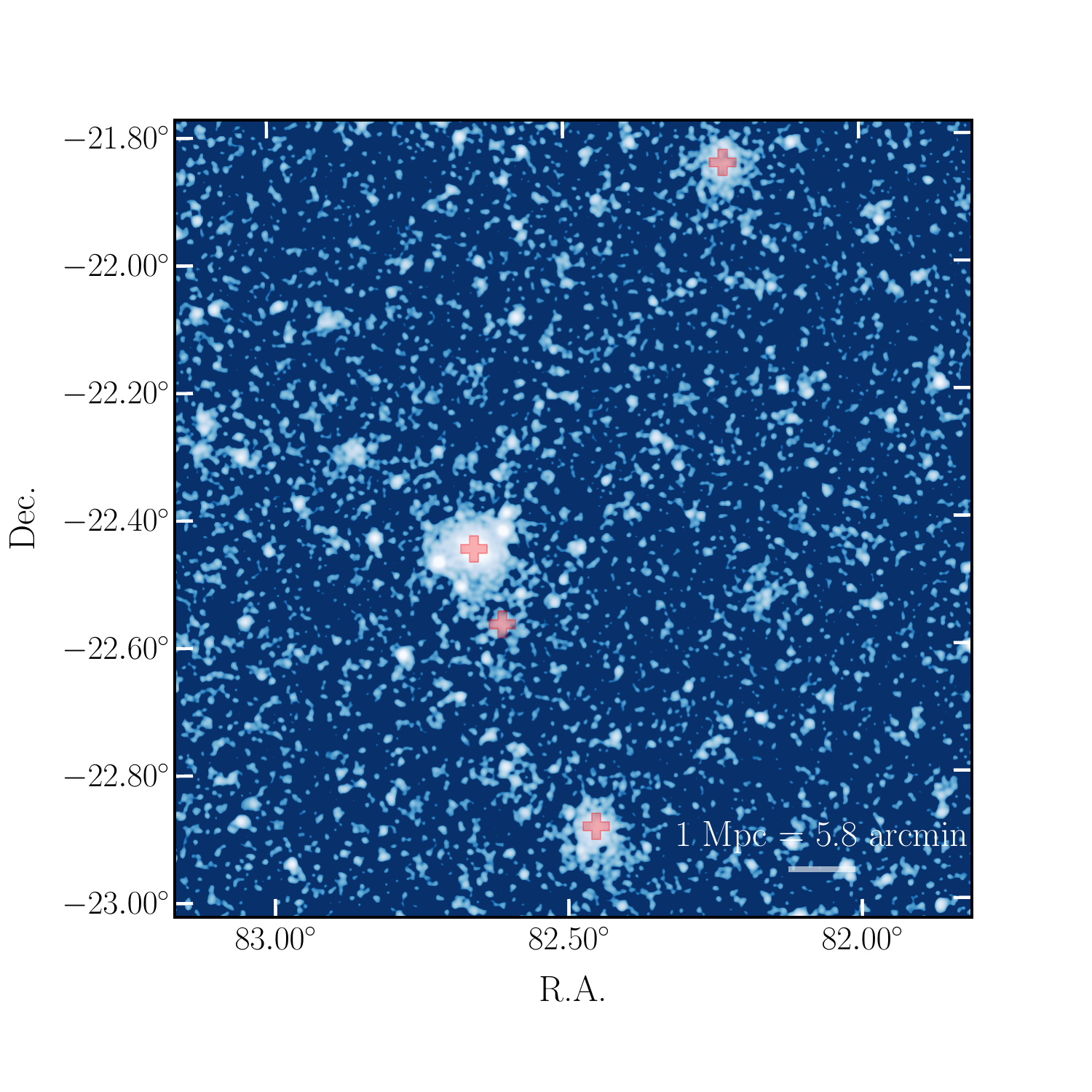}
\includegraphics[width=0.495\textwidth, trim=0 40 40 60, clip]{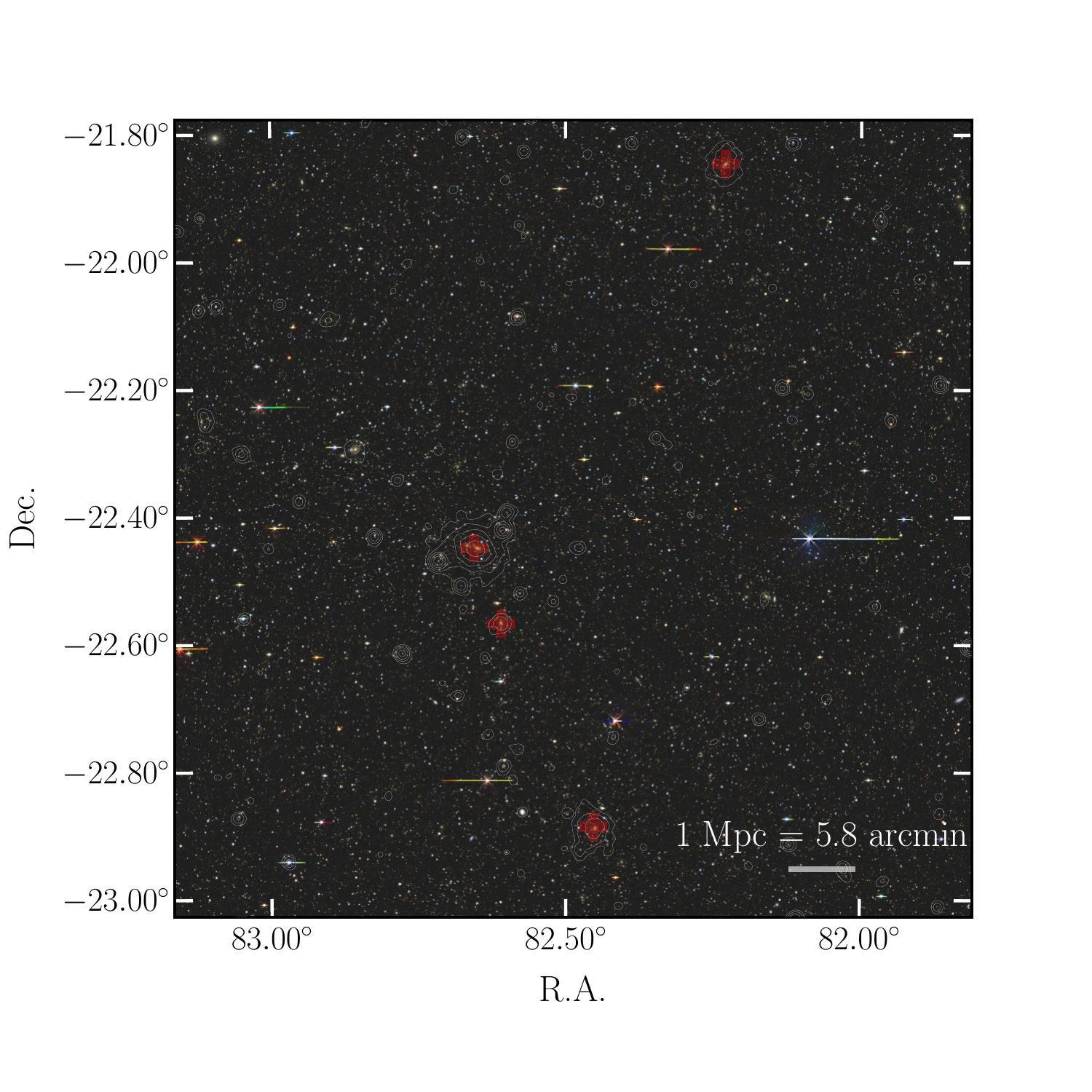}
\caption{\label{fig:opt} Examples of supercluster systems detected in eRASS1. Upper panels: the most distant supercluster system identified in eRASS1: 1eRASS-SC~J0530-4138, a cluster pair at redshift 0.802. The red crosses mark the position of the two member clusters: 1eRASS~J052957.4-413822 at $z=0.811\pm0.007$ and 1eRASS~J053040.8-413904 at $z=0.793\pm0.012$. Lower panels: a rich supercluster 1eRASS-SC~J0529-2226 detected at redshift 0.17, with the four member clusters marked as red crosses. The left panels show the {\sl eROSITA} X-ray exposure-corrected images in the 0.2--2.3~keV band, after smoothing with a Gaussian of $\sigma=12\arcsec$. The right panels show the optical images from the Legacy Survey, with the X-ray emission overlaid as white contours.  }
\end{center}
\end{figure*}

\begin{figure*}
\begin{center}
\includegraphics[width=0.99\textwidth, trim=70 15 20 10, clip]{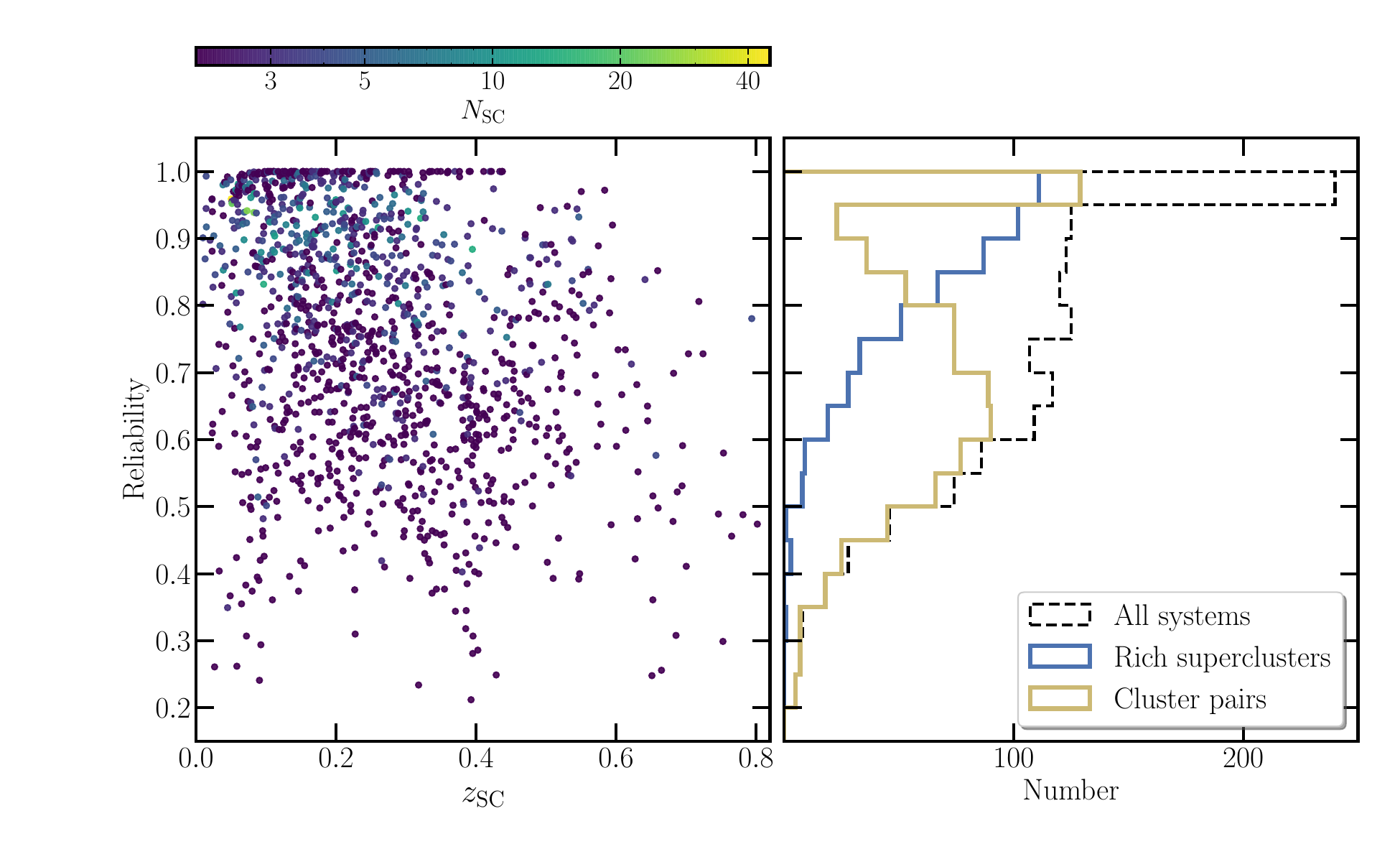}
\caption{\label{fig:robust} Reliability of the eRASS1 superclusters accounting for cluster redshift uncertainties. In the left panel, each data point represents a supercluster system color-coded by its multiplicity. Shown in the right panel is the histogram of reliability for all the systems (black), superclusters with $\ge 3$ members (blue), and cluster pairs (yellow). }
\end{center}
\end{figure*}

To identify superclusters in eRASS1, we adopt the similar FoF method as \citet{2022Liu}. For a random cluster in the sample, the algorithm searches for neighboring clusters (namely, ``friends'') closer than a specific distance, called the ``linking length'' $l$. Then, for each of the ``friends'', the algorithm continues to search for ``friends of friends'' until no new neighboring clusters are found. The co-moving distances between clusters are computed using their redshifts and X-ray centroids ({\tt RA\_XFIT, DEC\_XFIT}) obtained from the fit of {\tt MBProj2D}. 
The 3D comoving distance between two clusters, assuming a flat universe with $\Omega_k=0$, can be computed as:
\begin{equation}
\label{eqd1}
       D = \sqrt{D_{\rm C,1}^2+D_{\rm C,2}^2-2D_{\rm C,1} D_{\rm C,2} \cdot {\rm cos}\theta},
\end{equation}
where $D_{\rm C,1}$ and $D_{\rm C,2}$ are the line-of-sight co-moving distances of the two clusters at (R.A.=$\alpha_1$, Dec.=$\delta_1$, $z=z_1$) and (R.A.=$\alpha_2$, Dec.=$\delta_2$, $z=z_2$), $\theta$ is the angular separation of the clusters.
The line-of-sight co-moving distance of a cluster at redshift $z$ can be written as:
\begin{equation}
\label{eqd2}
       D_{\rm C}(z) = D_{\rm H}\cdot \int_{0}^{z}\frac{{\rm d}z'}{E(z')},
\end{equation}
where $D_{\rm H}\equiv c/H_0$ is the Hubble distance, and the function $E(z)$ is defined as:
\begin{equation}
\label{eqd3}
       E(z)\equiv \sqrt{\Omega_M (1+z)^3 + \Omega_k (1+z)^2 + \Omega_{\Lambda}} \ \ (\Omega_k=0).
\end{equation}
The angular separation of the two clusters $\theta$ can be computed as:
\begin{equation}
\label{eqd4}
       {\rm cos}\theta = {\rm sin}\delta_1 {\rm sin}\delta_2+{\rm cos}\delta_1 {\rm cos}\delta_2 {\rm cos}(\alpha_1-\alpha_2).
\end{equation}

The linking length, in co-moving 3D distance, is computed as a function of the average cluster number density $n$ and the desired overdensity ratio $f$: $l\equiv (n \times f)^{-1/3}$. Here, $n^{-1/3}$ is simply the average distance between two neighboring clusters assuming that the clusters are uniformly distributed in a 3D space without structures such as superclusters. Thus multiplying $n$ with the overdensity ratio $f$ implies that the cluster density in a supercluster is $f$ times that of the space average. In some other works, a factor of $4\pi/3$ is included in the computation of local cluster density, by assuming spherical collapse, so that the linking length becomes $l=(4\pi nf/3)^{-1/3}$ \citep[e.g.,][]{1993Zucca}. However, we note that most supercluster systems, especially cluster pairs, do not have regular shapes. Estimating the volume of a low-multiplicity supercluster with a sphere of radius $l$ will likely overestimate the volume and underestimate the density. We therefore use the more general definition of $l$ without the assumption of spherical collapse: $(n \times f)^{-1/3}$. 

The selection of eRASS1 clusters is a strong function of X-ray count rate, which confines the detection of high-redshift clusters to the high-luminosity regime, known as the Malmquist bias, a common selection bias for flux-limited surveys. Additionally, surface brightness patterns also have non-negligible effects on the selection of X-ray clusters: high redshift clusters with bright cores are more likely to be misidentified as point sources because of their smaller angular sizes and the large PSF of X-ray telescopes \citep[e.g.,][]{2022Bulbul}. The combination of these selection effects leads to a decrease in the number density of eRASS1 clusters, and thus a rapid increase in linking length, at high redshifts. Additionally, as the survey depth of eRASS1 is not uniform, where the ecliptic pole areas have longer exposure than the equatorial areas, the distribution of eRASS1 clusters generally shows the same pattern of nonuniformity as the eRASS1 exposure map. The magnitude of this nonuniformity is much lower compared to the evolution of cluster number density with redshift. However, for some studies on superclusters, for example, the comparison between supercluster members and isolated clusters, one would prefer a more uniform selection after accounting for the nonuniformity of the parent cluster sample. Therefore, in the computation of cluster number density, we consider not only the dependency on redshift but also the influence of exposure time. This is different than what \citet{2022Liu} have done on eFEDS clusters, where cluster density is computed only as a function of redshift because the survey depth of eFEDS is nearly uniform. Our linking length is then defined as:
\begin{equation}
\label{eqlk}
       l(z,t) = \left(\frac{N(z,t)}{V(z,t,A(t))}\cdot f\right)^{-1/3},
\end{equation}
where $N$ and $V$ are the number of clusters and the corresponding survey volume in the redshift range [$z-\delta z, z+\delta z$] and exposure range [$t-\delta t, t+\delta t$]. The thickness of the shell, $2\delta z$, is fixed at 0.01. $\delta t$ is adjusted to make sure the corresponding survey area is larger than 100 deg$^2$. $A(t)$ is the eRASS1 depth curve \citep{Bulbul2023}. $f$ is the overdensity ratio, and we adopt $f=10$, which is a common choice in many previous works \citep[e.g.,][]{2013Chon,2018Adami,2022Liu}. 

Since our cluster density decreases rapidly at high redshifts, we empirically set an upper limit on linking length: $l<=50$~Mpc, to avoid spurious detections at high redshifts. Namely, we set 50~Mpc as the maximum distance between two clusters that are believed to be connected with each other. We note that the choice of 50~Mpc as the upper limit of linking length is relatively conservative compared to previous works. This will help reduce false detections and systems with very large distances between members. As a comparison, the linking length can be as large as 70--80~Mpc in XXL \citep{2018Adami} and eFEDS \citep{2022Liu}. We also note that the thickness of the volume shell for computing the local cluster density in Eq.~\ref{eqlk}, $2\delta_z = 0.01$, ranges between 27~Mpc (at $z=0.8$) and 42~Mpc (at $z=0.05$), both are smaller than the upper limit of linking length.
We plot in Fig.~\ref{lk} the cluster density and linking length as functions of redshift and exposure time. As expected, the linking length shows a much stronger dependence on redshift than the exposure time. In summary, our selection of superclusters includes the cluster sample selection described in Sect.~\ref{sec:cluster2}, and the linking length plotted in the right panel of Fig.~\ref{lk}. These selection procedures need to be accounted for in statistical studies of superclusters and in the comparison with numerical simulations.

In the identification of superclusters, we use the best redshifts of the eRASS1 clusters ({\tt BEST\_Z} column in the eRASS1 cluster catalog) to compute the distance between clusters. However, the 3D distance between clusters is sensitive to the precision of cluster redshifts. Therefore, for each supercluster system detected using {\tt BEST\_Z}, we must check how robust the detection is over the clusters' redshift uncertainties. In addition to that, the peculiar velocities of clusters can also contribute to their redshifts and are also an important source of uncertainties in the distances. For clusters with spectroscopic redshifts, the distance uncertainties might be dominated by peculiar motions. The peculiar velocities of clusters are not well-constrained, except for the very nearby Universe. We, therefore, adopt a typical value of $v_{\rm pec} = cz_{\rm pec} = 400$~km/s \citep[see, e.g.,][]{2013Dolag} and add this component to the uncertainties of redshift in quadrature with the original redshift uncertainties. 
We then perform the following simulations to estimate the reliability of each supercluster over the total uncertainties of the clusters' redshifts. In each simulation, we randomly vary the redshift of each cluster in the sample over its total uncertainty. The same FoF method is then employed to identify superclusters in each simulated cluster sample. The simulations are performed 1000 times. We note that directly comparing the superclusters in each simulation by simply matching the central coordinates is not straightforward, and might bring misleading results. For instance, the addition or subtraction of a single member in the outskirts of a rich supercluster can cause a significant shift in its central position, even if the dominant part of the superclusters is robustly detected in both cases. Therefore, instead of estimating the detection rate of the superclusters in the simulations, we compute for each cluster the frequency of being a supercluster member. Then, for each supercluster detected using {\tt BEST\_Z}, we define its reliability as the average frequency of its member clusters.

Another important point to note is that our supercluster identification, similar to all the other supercluster catalogs in both optical and X-ray bands, is clearly affected by the selection of the parent eRASS1 cluster sample and the detection algorithm. In principle, there is no universal division between supercluster members and isolated clusters. Obviously, an isolated cluster identified in a shallower survey can become a supercluster member in a deeper survey, when its faint neighbors are detected. Therefore, one has to specify the selection of the cluster sample and the detection criteria associated with a supercluster sample. Similarly, the comparison between supercluster members and isolated clusters also requires that the two classes of clusters have consistent selection functions, despite that the division of the two classes might differ in different surveys.

With the 8862 eRASS1 clusters and the linking length defined in Eq.~\ref{eqlk}, we identify 1338 supercluster systems in eRASS1, including 818 cluster pairs and 520 rich superclusters with $>=3$ members. 3948 clusters are identified as members of these supercluster systems.
In addition to this primary supercluster catalog obtained with $f=10$, we also employ the same FoF method
to search for supercluster systems with a lower overdensity ratio $f=3$ and a higher overdensity ratio $f=50$. As a result, we detect 1270 and 929 superclusters corresponding to $f=3$ and $f=50$, respectively. In both cases, the numbers of superclusters are lower than that of $f=10$. This result is consistent with \citet{2013Chon} where the authors show in their Figure~2 that the number of detected superclusters reaches its maximum value between $f=5$ and $f=10$. We also note that 4867 and 2205 clusters are identified as supercluster members for $f=3$ and $f=50$, consistent with the trend that the lower the overdensity ratio, the more clusters are linked as superclusters. We base our following analysis on the primary supercluster catalog with $f=10$.

\begin{table*}
\caption{\label{tab:sc} General properties of superclusters identified in eRASS1.  }
\begin{center}
\begin{tabular}[width=\textwidth]{ccccccccc}
\hline\hline
Name & Redshift & RA & Dec & Multiplicity & Total mass & Reliability & $d_{\rm mem}$ & $L$ \\
\hline
[-] & [-] & [deg] & [deg] & [-] & [$10^{14} M_{\odot}$] & [-] &  [Mpc] & [Mpc]\\
\hline
... &  &  &  &  &  &  &  &  \\
1eRASS-SC J1208-8349 & 0.169 & 182.0510 & -83.8305 & 3 & 27.52$\pm$5.88 & 0.78 & 10.95 & 26.00 \\
1eRASS-SC J1338-0413 & 0.169 & 204.6015 & -4.2213 & 2 & 24.65$\pm$5.24 & 0.66 & 11.16 & 22.31 \\
1eRASS-SC J0527-4654 & 0.170 & 81.7702 & -46.9022 & 2 & 5.60$\pm$1.35 & 0.76 & 3.66 & 7.32 \\
1eRASS-SC J0442-5600 & 0.170 & 70.6324 & -56.0091 & 2 & 3.24$\pm$0.89 & 0.64 & 7.23 & 14.46 \\
1eRASS-SC J0349-3207 & 0.170 & 57.4379 & -32.1185 & 11 & 80.64$\pm$16.23 & 0.95 & 22.14 & 74.55 \\
... &  &  &  &  &  &  &  &  \\
\hline
\end{tabular}
\tablefoot{Column 1: supercluster name. Column 2: supercluster redshift, computed from the average of its member clusters. Columns 3 and 4: central coordinate of supercluster. Column 5: multiplicity, defined as the number of member clusters. Column 6: the total mass of superclusters, scaled from the masses of the member clusters, see the text for more details. Column 7: reliability. Column 8: average distance of supercluster members to the center. Column 9: supercluster length, defined as the largest distance between its members. The table is sorted by supercluster redshifts. Only part of the table is presented here. The full table is available at the CDS via anonymous ftp to \url{cdsarc.u-strasbg.fr} (130.79.128.5) or via \url{http://cdsweb.u-strasbg.fr/cgi-bin/qcat?J/A+A/}. }
\end{center}
\end{table*}

\begin{table*}
\caption{\label{tab:member} Properties of member clusters of the eRASS1 superclusters. }
\begin{center}
\begin{small}
\begin{tabular}[width=\textwidth]{cccccccc}
\hline\hline
SC name & Cluster name & Redshift & RA & Dec & $M_{500}$ & $L_{500}$ & $P$ \\
\hline
[-] & [-] & [-] & [deg] & [deg] & [$10^{14} M_{\odot}$] & [$10^{43}$ erg s$^{-1}$] & [-] \\
\hline
... &  &  &  &  &  &  &  \\
1eRASS-SC J1208-8349 & 1eRASS J121629.5-845559 & 0.1712$\pm$0.0050 & 184.1226 & -84.9345 & $1.25\pm0.43$ & $2.11\pm0.94$ & 0.66 \\
1eRASS-SC J1208-8349 & 1eRASS J121825.5-824718 & 0.1682$\pm$0.0044 & 184.6240 & -82.7890 & $2.06\pm0.25$ & $4.02\pm0.62$ & 0.83 \\
1eRASS-SC J1208-8349 & 1eRASS J114936.9-834601 & 0.1666$\pm$0.0043 & 177.4063 & -83.7681 & $4.04\pm0.34$ & $10.28\pm0.86$ & 0.84 \\
1eRASS-SC J1338-0413 & 1eRASS J134048.2-045446 & 0.1661$\pm$0.0004 & 205.2020 & -4.9161 & $1.55\pm0.34$ & $2.93\pm0.61$ & 0.65 \\
1eRASS-SC J1338-0413 & 1eRASS J133600.2-033133 & 0.1726$\pm$0.0004 & 204.0011 & -3.5265 & $5.04\pm0.39$ & $16.31\pm1.47$ & 0.67 \\
1eRASS-SC J0527-4654 & 1eRASS J052823.3-464208 & 0.1673$\pm$0.0004 & 82.0974 & -46.7015 & $1.02\pm0.15$ & $1.55\pm0.32$ & 0.77 \\
1eRASS-SC J0527-4654 & 1eRASS J052548.3-470619 & 0.1719$\pm$0.0043 & 81.4431 & -47.1028 & $0.48\pm0.15$ & $0.60\pm0.20$ & 0.75 \\
... &  &  &  &  &  &  & \\
\hline
\end{tabular}
\tablefoot{Column 1: supercluster name. Column 2: member cluster name. Column 3: cluster redshift. Columns 4 and 5: cluster coordinate. Column 6: $M_{500}$. Column 7: $L_{500}$. Column 8: membership probability. The table is sorted by supercluster redshifts. Only part of the table is presented here. The full table is available at the CDS via anonymous ftp to \url{cdsarc.u-strasbg.fr} (130.79.128.5) or via \url{http://cdsweb.u-strasbg.fr/cgi-bin/qcat?J/A+A/}. }
\end{small}
\end{center}
\end{table*}

\begin{figure}
\begin{center}
\includegraphics[width=0.495\textwidth, trim=85 45 70 50, clip]{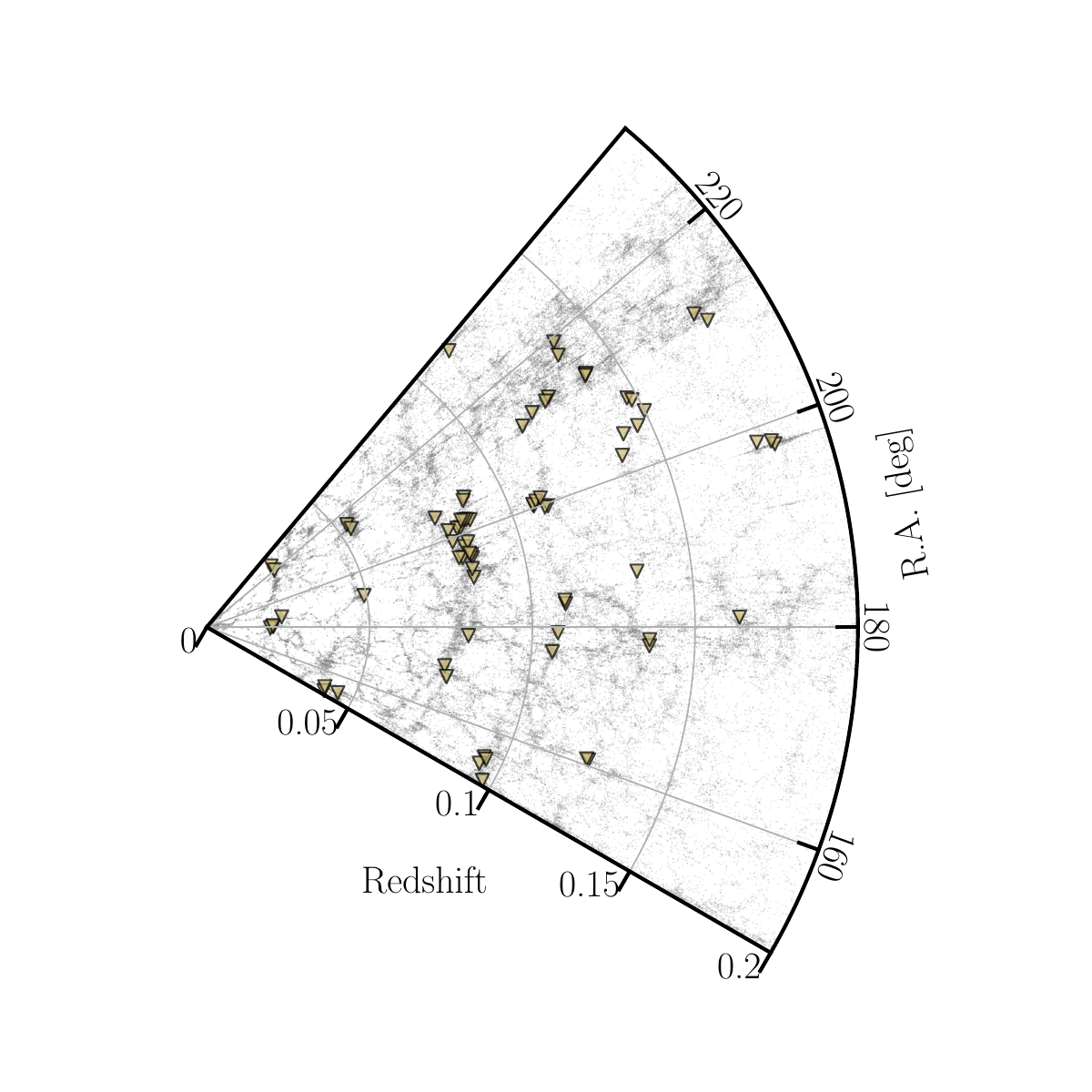}
\caption{\label{fig:lss} Comparison of the large-scale structures traced by galaxies (gray dots) and X-ray superclusters identified in this work. The redshifts and positions of the galaxies are from a spectroscopic galaxy compilation of published catalogs \citep[see][]{Kluge2023}. Each yellow triangle represents a supercluster member. Plotted is the slice $-3^{\circ} <$ Dec. $< 3^{\circ}$.  }
\end{center}
\end{figure}

\section{Properties of the eRASS1 superclusters}
\label{sec:property}

We present in this section the properties of the primary eRASS1 supercluster catalog. The spatial distribution of the rich superclusters is presented in Fig.~\ref{fig:scall}. Although we have accounted for the dependency of cluster density on exposure time, there is still a slight overdensity of identified supercluster systems around the South Ecliptic Pole, where eRASS1 is deeper than other regions. 
Among the 8862 clusters in the sample, 3948 ($45\%$) are identified as supercluster members. This fraction is only slightly lower than the result in \citet{Boehringer2021}, where $51\%$ of the clusters are found to be supercluster members. The difference is likely due to the fact that \citet{Boehringer2021} adopt a much lower overdensity ratio ($f=2$). The distribution of multiplicity, defined as the number of member clusters in a supercluster, is shown in Fig.~\ref{fig:multi}. The richest system we identify, 1eRASS-SC~J1307-3016, also known as the Shapley supercluster, has 45 members with a median redshift of 0.050. 
For each supercluster system, we compute the average distance of its members to the center of the system ($d_{\rm mem}$, where the center is defined as the algebraic mean coordinate of the members), and the total length ($L$), which is defined as the maximum distance between its members. Both quantities are computed in the projected 2D distance. The distribution of the average separation and length are shown in Fig.~\ref{fig:dist}. 
$d_{\rm mem}$ can be used to quantify how compact is a supercluster system. For rich superclusters, the distribution of $d_{\rm mem}$ reaches a peak value at around $15$~Mpc, and extends to 40~Mpc (see the left panel of Fig.~\ref{fig:dist}). The median $d_{\rm mem}$ of rich superclusters is 12.9~Mpc. On the other hand, most of the cluster pairs have a low $d_{\rm mem}$ smaller than $15$~Mpc, with a small fraction extending to 25~Mpc, due to the upper limit of 50~Mpc on linking length. The median $d_{\rm mem}$ of cluster pairs is 8.4~Mpc, much lower than rich superclusters. A similar trend can be found in the distribution of total length $L$ (see the right panel of Fig.~\ref{fig:dist}). The $L$ of rich superclusters peaks at around $30$~Mpc, with a median value of 33.5~Mpc, while the median $L$ of cluster pairs is 16.9~Mpc.
The most extensive system we detect in eRASS1, 1eRASS-SC~J1140-1939 at redshift 0.303, consisting of 10 member clusters, has a projected comoving length of 127~Mpc. As a comparison, the superclusters identified with SDSS DR7 data \citep{2012Liivamagi} have average and maximum diameters of 22~Mpc and 120~Mpc, similar to the sizes of the eRASS1 X-ray superclusters. 

We also estimate the total cluster mass of the supercluster systems by simply summing up the virial masses $M_{\rm 200}$ of its member clusters, where we adopt the approximation $M_{\rm 200}\approx 1.46\times M_{500}$ by assuming a Navarro-Ferenk-White (NFW) profile \citep{1997Navarro} with concentration $c\equiv r_{200}/r_{\rm s}=4$ \citep{2013Reiprich}. The total cluster mass is then converted to the total supercluster mass by adopting the relation found by \citet{2014Chon} with cosmological $N$-body simulations: $M_{\rm tot,Cl} = 0.39\pm0.077 \times M_{\rm tot,SC}$. The most massive eRASS1 supercluster is the Shapley supercluster, 1eRASS-SC~J1307-3016, with a total mass of $2.58\pm0.51\times 10^{16} M_{\odot}$ and a length of $L=111$~Mpc. The mass estimation of the Shapley supercluster is available in several previous works. For example, \citet{2000Reisenegger} reported a total mass of $1.9\times 10^{16} M_{\odot}$ within $12$~Mpc, using a caustic method of galaxies, and assuming a spherical collapse model. \citet{2006Ragone} measured a total mass of $2.3\times 10^{16} M_{\odot}$, using 122 galaxy systems in a area of $12\times15$ deg$^2$, with the masses outside the galaxy systems corrected. More recently, \citet{2014Chon} found a total mass of $1.91\times 10^{16} M_{\odot}$ within $17.7$~Mpc. We note that, given the difference in radius, data, and methods, our result is in broad agreement with the values reported by the previous works. A dedicated analysis on the Shapley supercluster with {\sl eROSITA} will be performed in another work (Sanders et al., in prep.).

\begin{figure*}
\begin{center}
\includegraphics[width=\textwidth, trim=10 10 110 10, clip]{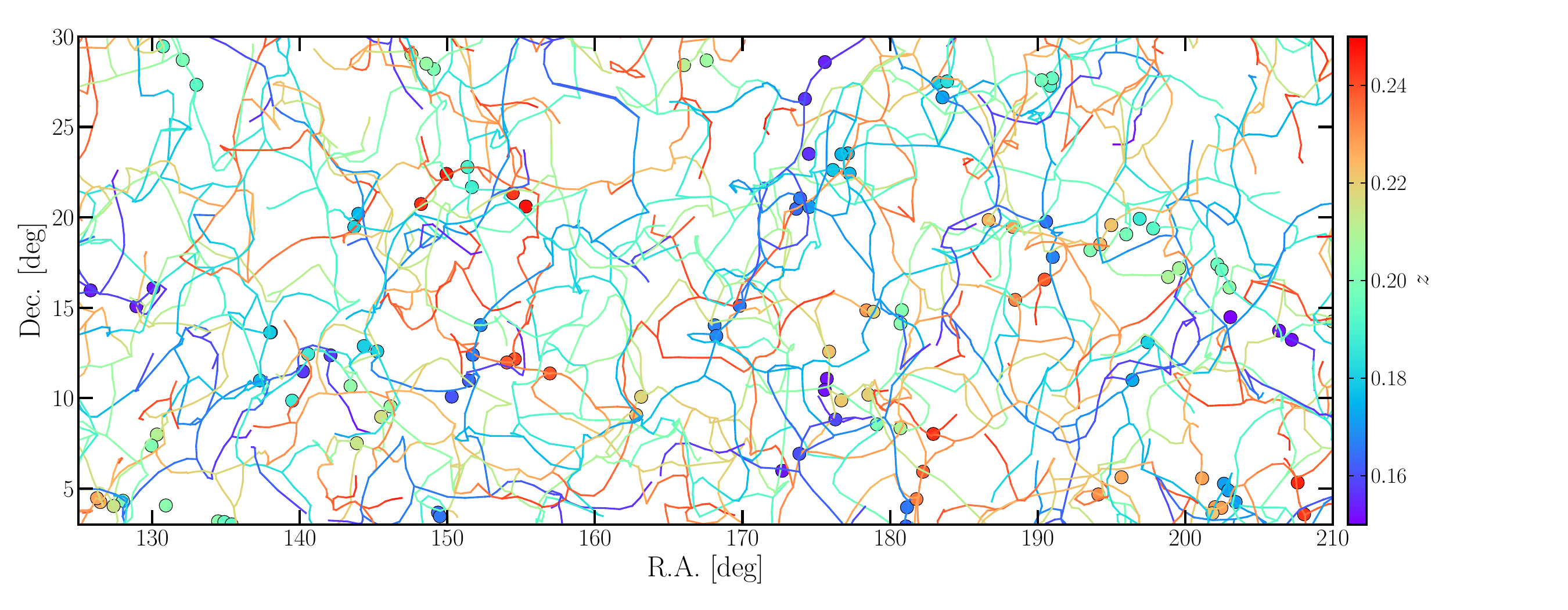}
\caption{\label{fig:fi} Comparison of the filaments detected with SDSS galaxies \citep{2020Malavasi} (curves) and the eRASS1 X-ray superclusters. The supercluster members are plotted with dots. Color codes denote redshifts for both the filaments and supercluster members. To reduce overlapping between optical filaments, only the region ($125^\circ <$ R.A. $< 210^\circ$, $3^\circ < $ Dec. $< 30^\circ$) and redshift range $0.15 < z < 0.25$ are plotted in the figure. }
\end{center}
\end{figure*}

The redshift distribution of the supercluster systems is shown in Fig.~\ref{fig:scz}. The supercluster systems at high redshifts are dominated by cluster pairs, owing to the decrease in the number density of clusters in the sample. Also shown in Fig.~\ref{fig:scz} in the right panel is the mass-redshift relation of rich superclusters with $\ge 3$ members, where the masses of the systems span a range of [$6\times 10^{13}$--2$\times$ 10$^{16} M_{\odot}$].
The most distant system, 1eRASS-SC~J0530-4138, is a cluster pair at redshift $0.802$, consisting of two members, 1eRASS~J052957.4-413822 and 1eRASS~J053040.8-413904. We show in Fig.~\ref{fig:opt} the optical image of this cluster pair from the Legacy Imaging Survey. The photometric redshifts of the two members are $0.811\pm0.007$ and $0.793\pm0.012$, and the masses ($M_{500}$) are $3.6_{-0.8}^{+1.0}\times 10^{14} M_{\odot}$ and $4.2_{-0.6}^{+0.8}\times 10^{14} M_{\odot}$. Due to the large error bar in redshifts, we are not able to precisely constrain the 3D distance between the two members. Therefore, this system has relatively low reliability, $P=0.50$. The projected distance is about $3.7$~Mpc, implying that the two members will probably merge as a massive cluster. Although this cluster pair is identified for the first time in this work, the two member clusters are already detected in Sunyaev–Zeldovich (SZ) surveys by the Atacama Cosmology Telescope \citep[ACT,][]{2021Hilton} and the South Pole Telescope \citep[SPT,][]{2019Bocquet}. The reported photometric redshifts of 1eRASS~J052957.4-413822 and 1eRASS~J053040.8-413904 are $0.793\pm0.010$ and $0.795\pm0.010$ in ACT, and $0.775\pm0.050$ and $0.775\pm0.048$ in SPT. Therefore, our redshifts are consistent with both SZ surveys within $2\sigma$, indicating that the detection of this cluster pair is reliable.
Also shown in Fig.~\ref{fig:opt} is another example of a rich eRASS1 supercluster: 1eRASS-SC~J0529-2226 consisting of four members at an average redshift of 0.17.

The reliabilities of the supercluster systems are computed using the method described in Sect.~\ref{sec:detection}, and are presented in Fig.~\ref{fig:robust}. Among the 1338 systems, 841 ($63\%$) have reliability larger than 0.7. For rich superclusters with $\ge3$ members and cluster pairs, this fraction is $87\%$ and $48\%$, indicating that the identification of rich superclusters is generally more reliable than cluster pairs. On average, low-redshift systems have higher reliability than high-redshift ones, as can be seen from the left panel of Fig.~\ref{fig:robust}. This is probably due to the fact that the fraction of spectroscopic redshift is larger at low redshifts (see Fig.~\ref{sample}), which is about $10\times$ more precise than photometric redshift. Dedicated spectroscopic follow-up of the {\sl eROSITA} clusters is ongoing or planned with SDSS-V \citep{2017Kollmeier,2023Almeida} and 4MOST \citep{2019Finoguenov}. It is expected that the reliability of the superclusters will be further improved when more spectroscopic redshifts are available.

The properties of the eRASS1 superclusters, including the average redshift, coordinate, multiplicity, total mass, average member separation $d_{\rm mem}$, and total length $L$, are present in Table~\ref{tab:sc}. Since most of the superclusters do not have a regular shape, the average redshift and coordinate alone cannot locate the supercluster precisely. We, therefore, provide in Table~\ref{tab:member} the properties of the member clusters for all the superclusters. 
In addition to the primary supercluster catalog obtained with $f=10$, we also provide in the appendix a supplementary supercluster catalog with the information of members, corresponding to $f=50$.

We compare our eRASS1 supercluster catalog with the known superclusters published in the literature. In many of the published supercluster catalogs, the location of superclusters is given as an average coordinate of the members. As superclusters are not bounded systems with irregular shapes, and they often span an area of several square degrees, it is almost infeasible to directly match the superclusters with the central coordinates. A practical way is to match the member clusters. 

We first compare the eRASS1 supercluster catalog with eFEDS. As eFEDS is almost $10$ times deeper than eRASS1, only a small fraction (63 out of 542) of eFEDS clusters are detected in eRASS1. In the eFEDS survey, \citet{2022Liu} detected 19 rich superclusters with $\ge 4$ members. In this work, using the eRASS1 cluster sample, we detect 8 superclusters (including cluster pairs) in the eFEDS footprint. Most of them are already found in the eFEDS survey, despite that we report lower multiplicities for some systems in this work than the values in \citet{2022Liu} due to the much lower source density in eRASS1 compared to eFEDS. For example, eFEDS-SC3 ($z=0.196, N=10$) is identified as 1eRASS-SC~J0859+0306 ($z=0.197, N=3$) in this work; eFEDS-SC5 ($z=0.269, N=7$) is identified as 1eRASS-SC~J0841-0036 ($z=0.267, N=2$) in this work; eFEDS-SC6 ($z=0.281, N=4$) is identified as 1eRASS-SC~J0921+0221 ($z=0.283, N=2$) in this work; eFEDS-SC12 ($z=0.358, N=4$) is identified as 1eRASS-SC~J0935+0051 ($z=0.359, N=2$) in this work. The only exception is 1eRASS-SC~J0932-0110, a cluster pair at $z=0.238$. The two members of this system both have counterparts in the eFEDS cluster sample. However, one of its members, 1eRASS~J093024.6-020635 (eFEDS~J093025.7-020507), has a lower redshift in the eFEDS cluster catalog: 0.220$\pm 0.006$, while the redshift measured in the eRASS1 cluster catalog is 0.241$\pm 0.006$. Both results are photometric redshifts. Therefore, it is not identified as a cluster pair in eFEDS. We note that the optical confirmation of eFEDS clusters was performed with the multi-component matched filter cluster confirmation tool \citep[MCMF,][]{2022Klein} and the data from HSC-SSP, Legacy Survey DR8, and unWISE. In eRASS1, we use the {\tt eROMaPPer} tool on Legacy Survey DR9 and DR10 \citep{Kluge2023}. The slight difference in the redshift of this cluster is probably because we use different optical survey data and methods for cluster confirmation in eFEDS and eRASS1.

35 rich supercluster systems and 39 cluster pairs are detected using the XXL365 cluster catalog \citep{2018Adami}. In eRASS1, we only detect 11 XXL clusters \citep{Bulbul2023} due to the relatively shallow depth of eRASS1 compared to XXL, thus consequently limiting the detection rate of XXL superclusters. Only one cluster pair is identified in eRASS1: Id17 in the XXL cluster pair catalog at redshift 0.378. In eRASS1, we identify this cluster pair as a rich system: 1eRASS-SC~J2321-5326, with four member clusters at a median redshift of 0.369. Except for the two members reported in \citet{2018Adami}, XLSSC513 and XLSSC525, we find two additional members: 1eRASS~J232541.0-531638 at $z=0.370$ (XLSSC547) and 1eRASS~J232708.1-513733 at $z=0.364$. 

We also make a general comparison with large-area optical spectroscopic surveys. In Fig.~\ref{fig:lss}, we plot a slice of cosmic volume at $z<0.2$ and $-3^{\circ} < $ Dec. $ < 3^{\circ}$. A few more examples for different slices are shown in Fig.~\ref{fig:lss2}. We find a broad association in the large-scale structures formed by galaxies and traced by the X-ray superclusters we identified in this work. We further compare the eRASS1 superclusters with the cosmic filaments detected with SDSS galaxies \citep{2020Malavasi} in Fig.~\ref{fig:fi}. To reduce overlapping between optical filaments, only a small part of the sky ($125^\circ < $R.A. $< 210^\circ$, $3^\circ < $ Dec. $< 30^\circ$) and redshift range $0.15 < z < 0.25$ are shown in Fig.~\ref{fig:fi}. As a result, most of the eRASS1 superclusters are associated with the filaments. Most of the supercluster members are connected by galaxy filaments. 

\section{Conclusions}
\label{sec:conclusions}
We present in this work the first catalog of superclusters in the western Galactic hemisphere detected from the eRASS1 survey. We base our supercluster detection on the optically-confirmed galaxy clusters detected in eRASS1 \citep{Bulbul2023}. By applying several additional filters on the primary eRASS1 cluster catalog, with the aim of obtaining a purer subsample of clusters with more reliable redshifts, we select a subsample of 8862 eRASS1 clusters. Based on a contamination estimator $P_{\rm cont}$ computed using the cluster's X-ray count rate, redshift, and optical richness, this cluster subsample has a high purity of 96.4\%. A Friends-of-Friends method is employed to identify supercluster systems, where the linking length as a function of redshift and exposure time is computed to make sure that the local cluster density in superclusters is at least 10 times that of the average value of the sample. The false detection rate at high redshifts where the cluster density is too low is controlled by conservatively setting an upper limit of 50~Mpc on the linking length.

With the above data and method, we identify 1338 supercluster systems up to redshift 0.8, including 818 cluster pairs and 520 rich superclusters with $\ge$ 3 members. Among the 8862 selected clusters, 3948 clusters (about 45\% of the sample) are members of these supercluster systems. The most massive and richest system, 1eRASS-SC~J1307-3016, also known as the Shapley supercluster, consists of 45 members at an average redshift 0.050, and has a total mass of $2.58\pm0.51\times10^{16} M_{\odot}$. The most extensive system, 1eRASS-SC~J1140-1939 at redshift 0.303, has a total length of 127~Mpc. The sizes of the superclusters we identify in this work are comparable to the structures found with galaxy survey data. A good association is found between the eRASS1 superclusters and the large-scale structures formed by optical galaxies.
We compute the reliability of each supercluster by accounting for the uncertainties in cluster redshifts and the peculiar velocities of the clusters. Thanks to the high accuracy of eRASS1 clusters' redshifts, 63\% of the supercluster systems have a reliability larger than 0.7. This will be further improved when more spectroscopic redshifts of the eRASS1 clusters are available from SDSS-V \citep{2017Kollmeier,2023Almeida} and 4MOST \citep{2019Finoguenov}.

The eRASS1 supercluster catalog presented in this work represents the most extensive sample of superclusters detected in the X-ray band in terms of sample volume, sky coverage, redshift range, and the availability of X-ray properties. Another advantage of the eRASS1 supercluster catalog is that the superclusters are identified on the basis of a cluster sample with a well-understood selection function, thus making it convenient to compare with large numerical simulations. This legacy catalog will greatly advance our understanding of the evolution of the cosmic large-scale structure. 

In a forthcoming paper (Liu et al. in prep.), we will utilize the eRASS1 supercluster catalog to investigate the environmental effects on the evolution of galaxy clusters. This can be done by simply comparing the X-ray properties of supercluster members and isolated clusters after accounting for selection effects \citep[see, e.g.,][]{2021Manolopoulou}.
As these two classes of clusters naturally have different clustering magnitudes, such a comparison can also be used to test the dark matter halo formation theories, such as the halo assembly bias \citep{Gao2005,Gao2007}. 

A task for X-ray astronomy in the next decade is to find the ``missing baryons'', which account for 30\%--40\% of the total baryonic mass in the Universe \citep[e.g.,][]{2007Bregman,2012Shull}. These ``missing baryons'' are probably residing in the vast space between the nodes of the cosmic web, namely, galaxies and clusters, in the form of a low-density and X-ray emitting gas, with a temperature of $\sim 10^6$~K \citep[e.g.,][]{2018Fang,2018Nicastro}. They can be detected either from their emission in the soft X-ray band \citep[e.g.,][]{2015Eckert,2016Bulbul,2023Veronica} or through their absorption on bright background sources \citep[][]{2022Nicastro,2023Stofanova}. 
Detecting and characterizing the intergalactic medium (IGM) and circumgalactic medium (CGM) are important science goals for X-ray missions in the near future, such as the Hot Universe Baryon Surveyor
 \citep[HUBS,][]{2023Bregman} and {\sl Athena} \citep{2013Nandra}. 
Superclusters are the most probable reservoirs of the missing baryons. Therefore, the eRASS1 X-ray supercluster catalog will provide promising targets for future X-ray missions. Limited by the survey depth of {\sl eROSITA}, the IGM emission in most of the individual supercluster systems is expected to be fainter than the detection limit of eRASS. Thus we need to apply image and spectrum stacking techniques to search for possible X-ray emissions from these cosmic filaments. The results of this work are presented in a separate paper \citep{Zhang2023}. Dedicated multiwavelength analysis of individual systems, for example, the Shapley supercluster, will also be performed, to investigate the merging processes in member clusters (Sanders et al. in prep., di Gennaro et al. in prep.). 

\begin{acknowledgement}
We greatly thank the anonymous referee for his/her constructive comments that helped improve the paper. 
This work is based on data from eROSITA, the soft X-ray instrument aboard SRG, a joint Russian-German science mission supported by the Russian Space Agency (Roskosmos), in the interests of the Russian Academy of Sciences represented by its Space Research Institute (IKI), and the Deutsches Zentrum für Luft- und Raumfahrt (DLR). The SRG spacecraft was built by Lavochkin Association (NPOL) and its subcontractors and is operated by NPOL with support from the Max Planck Institute for Extraterrestrial Physics (MPE).

The development and construction of the eROSITA X-ray instrument was led by MPE, with contributions from the Dr. Karl Remeis Observatory Bamberg \& ECAP (FAU Erlangen-Nuernberg), the University of Hamburg Observatory, the Leibniz Institute for Astrophysics Potsdam (AIP), and the Institute for Astronomy and Astrophysics of the University of Tübingen, with the support of DLR and the Max Planck Society. The Argelander Institute for Astronomy of the University of Bonn and the Ludwig Maximilians Universität Munich also participated in the science preparation for eROSITA.
\\
The eROSITA data shown here were processed using the eSASS/NRTA software system developed by the German eROSITA consortium.
\\
A. Liu, E. Bulbul, V. Ghirardini, C. Garrel, S. Zelmer, and X. Zhang acknowledge financial support from the European Research Council (ERC) Consolidator Grant under the European Union’s Horizon 2020 research and innovation program (grant agreement CoG DarkQuest No 101002585). M.B. acknowledges support from the DFG under Germany's Excellence Strategy - EXC 2121 ``Quantum Universe'' – 390833306.
\\
This work made use of SciPy \citep{jones_scipy_2001}, Matplotlib, a Python library for publication-quality graphics \citep{Hunter2007}, Astropy, a community-developed core Python package for Astronomy \citep{Astropy2013}, NumPy \citep{van2011numpy}. 

\end{acknowledgement}

\bibliography{main}


\begin{appendix}

\onecolumn

\section{Supplementary catalog of superclusters identified with $f = 50$ }

\begin{table*}
\caption{\label{tab:sc50} General properties of superclusters identified in eRASS1 with overdensity ratio $f = 50$. }
\begin{center}
\begin{tabular}[width=\textwidth]{ccccccccc}
\hline\hline
Name & Redshift & RA & Dec & Multiplicity & Total mass & Reliability & $d_{\rm mem}$ & $L$ \\
\hline
[-] & [-] & [deg] & [deg] & [-] & [$10^{14} M_{\odot}$] & [-] &  [Mpc] & [Mpc]\\
\hline
... &  &  &  &  &  &  &  &  \\
1eRASS-SC J0538-4823 & 0.196 & 84.6960 & -48.3927 & 2 & 6.87$\pm$1.77 & 0.45 & 1.57 & 3.13 \\
1eRASS-SC J1242+2731 & 0.196 & 190.7033 & 27.5222 & 3 & 38.61$\pm$8.07 & 0.89 & 4.51 & 8.93 \\
1eRASS-SC J0332-0711 & 0.196 & 53.0321 & -7.1875 & 2 & 15.67$\pm$4.89 & 0.70 & 6.62 & 13.25 \\
1eRASS-SC J0329-2056 & 0.196 & 52.2812 & -20.9459 & 2 & 33.81$\pm$7.10 & 0.39 & 6.51 & 13.01 \\
1eRASS-SC J1818-7208 & 0.196 & 274.5161 & -72.1427 & 2 & 25.11$\pm$5.51 & 0.61 & 6.89 & 13.79 \\
... &  &  &  &  &  &  &  &  \\
\hline
\end{tabular}
\tablefoot{The meanings of columns are the same as Table~\ref{tab:sc}. Only part of the table is presented here. The full table is available at the CDS via anonymous ftp to \url{cdsarc.u-strasbg.fr} (130.79.128.5) or via \url{http://cdsweb.u-strasbg.fr/cgi-bin/qcat?J/A+A/}.}
\end{center}
\end{table*}

\begin{table*}
\caption{\label{tab:member50} Properties of member clusters of the eRASS1 superclusters identified with overdensity ratio $f = 50$. }
\begin{center}
\begin{small}
\begin{tabular}[width=\textwidth]{cccccccc}
\hline\hline
SC name & Cluster name & Redshift & RA & Dec & $M_{500}$ & $L_{500}$ & $P$ \\
\hline
[-] & [-] & [-] & [deg] & [deg] & [$10^{14} M_{\odot}$] & [$10^{43}$ erg s$^{-1}$] & [-] \\
\hline
... &  &  &  &  &  &  &  \\
1eRASS-SC J0538-4823 & 1eRASS J053835.4-481700 & 0.1935$\pm$0.0047 & 84.6509 & -48.2846 & $0.68\pm0.18$ & $1.11\pm0.34$ & 0.45 \\
1eRASS-SC J0538-4823 & 1eRASS J053857.4-482954 & 0.1978$\pm$0.0052 & 84.7412 & -48.5007 & $1.15\pm0.24$ & $2.19\pm0.55$ & 0.45 \\
1eRASS-SC J1242+2731 & 1eRASS J124325.7+271700 & 0.1924$\pm$0.0022 & 190.8563 & 27.2810 & $5.83\pm0.57$ & $20.96\pm2.19$ & 0.81 \\
1eRASS-SC J1242+2731 & 1eRASS J124356.3+274155 & 0.1958$\pm$0.0013 & 190.9820 & 27.6989 & $2.19\pm0.31$ & $5.04\pm0.94$ & 0.99 \\
1eRASS-SC J1242+2731 & 1eRASS J124104.0+273521 & 0.1990$\pm$0.0008 & 190.2715 & 27.5866 & $2.29\pm0.30$ & $5.77\pm0.98$ & 0.87 \\
... &  &  &  &  &  &  & \\
\hline
\end{tabular}
\tablefoot{The meanings of columns are the same as Table~\ref{tab:member}. Only part of the table is presented here. The full table is available at the CDS via anonymous ftp to \url{cdsarc.u-strasbg.fr} (130.79.128.5) or via \url{http://cdsweb.u-strasbg.fr/cgi-bin/qcat?J/A+A/}. }
\end{small}
\end{center}
\end{table*}

\section{A few more examples of comparison between eRASS1 superclusters and optical LSS }

\begin{figure*}
\begin{center}
\includegraphics[width=0.495\textwidth, trim=50 30 60 40, clip]{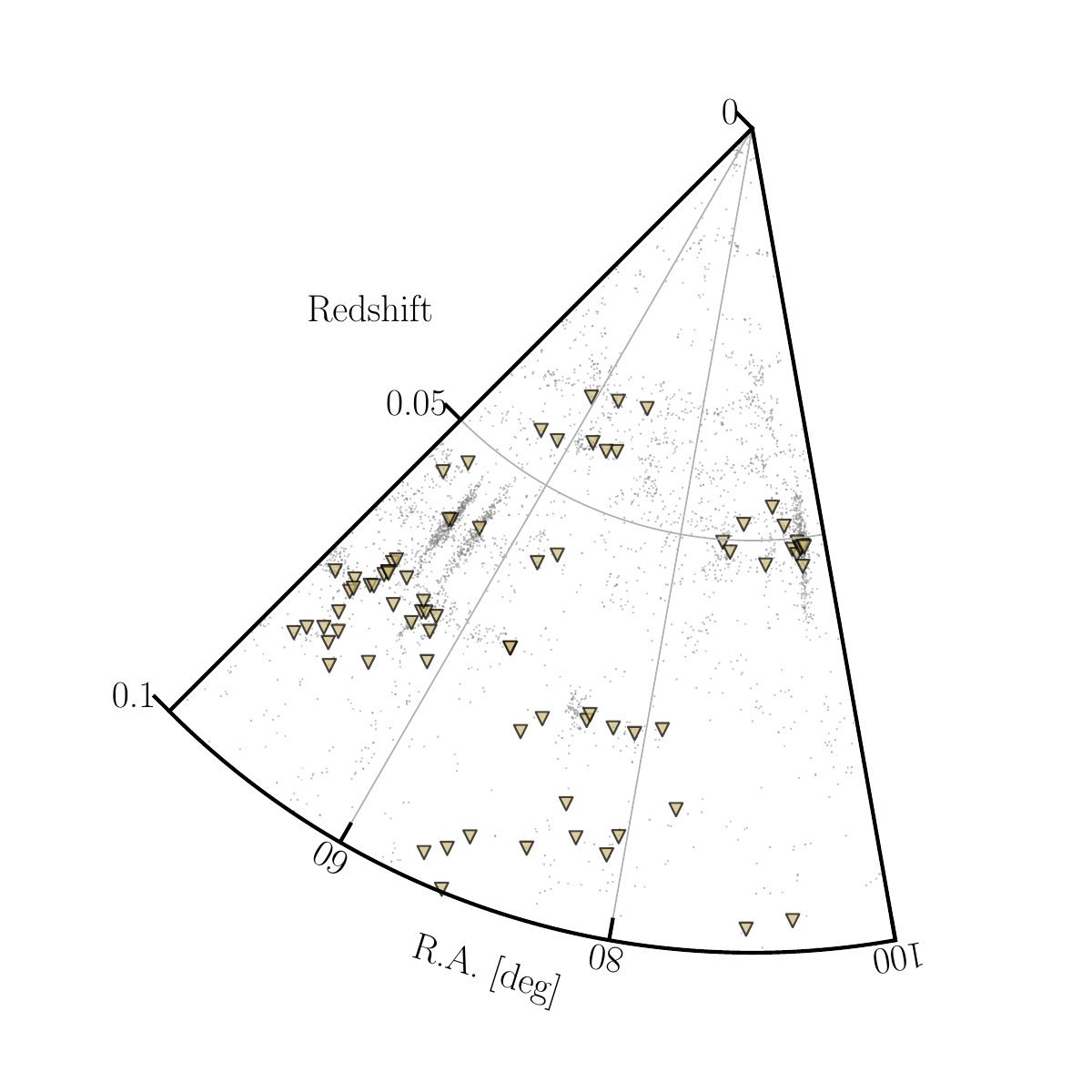}
\includegraphics[width=0.495\textwidth, trim=60 30 60 40, clip]{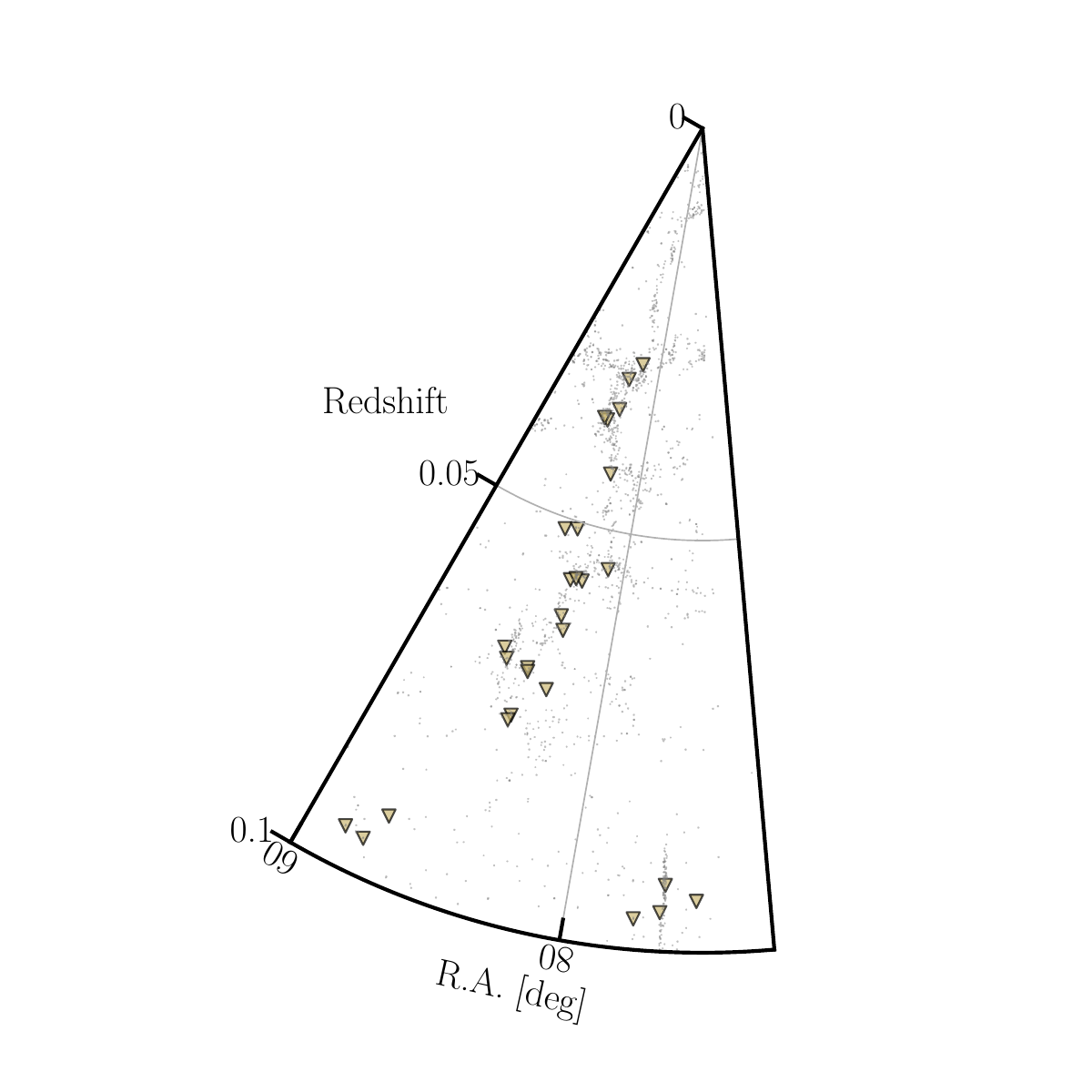}
\includegraphics[width=0.7\textwidth, trim=50 55 10 105, clip]{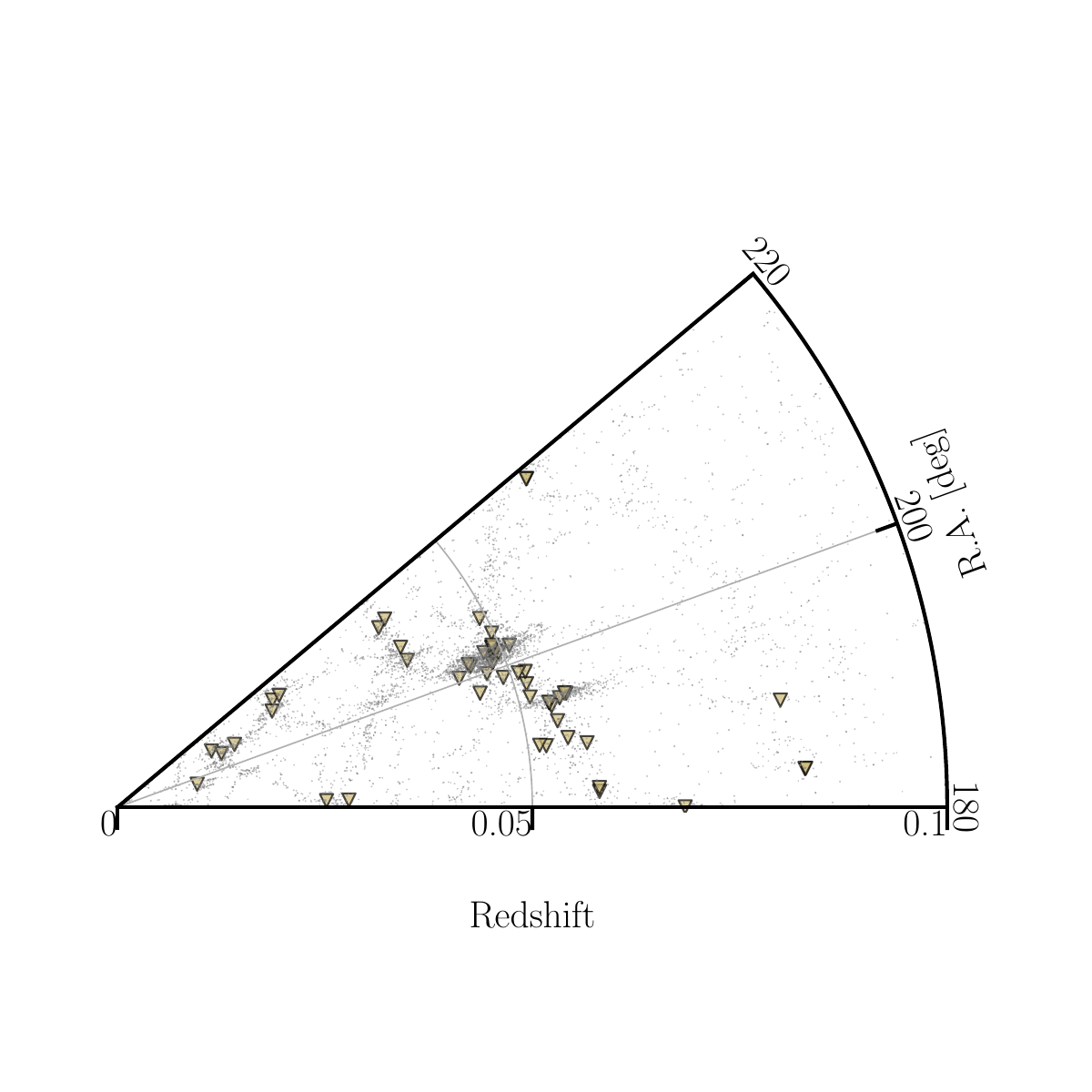}
\caption{\label{fig:lss2} Same as Fig.~\ref{fig:lss}, but for different slices. Top left: $45^{\circ} <$ R.A. $< 100^{\circ}$, $-55^{\circ} <$ Dec. $< -45^{\circ}$. Top right: $60^{\circ} <$ R.A. $< 95^{\circ}$, $-22^{\circ} <$ Dec. $< -15^{\circ}$. Bottom: $180^{\circ} <$ R.A. $< 220^{\circ}$, $-35^{\circ} <$ Dec. $< -25^{\circ}$. }
\end{center}
\end{figure*}

\end{appendix}

\end{document}